\documentclass[12pt]{article}
\pdfoutput=1
\usepackage[utf8]{inputenc}
\usepackage[DIV13]{typearea}
\usepackage{ragged2e}
\usepackage{calligra,amsmath,amsfonts,bbm,mathrsfs,amssymb}
\usepackage{slashed,cancel,units}
\usepackage[usenames,dvipsnames]{xcolor}
\usepackage{cite}
\usepackage{hyperref}
\hypersetup{colorlinks=true,urlcolor=Magenta,anchorcolor=blue,citecolor=blue,filecolor=blue,
            linkcolor=Magenta,menucolor=blue, linktocpage=true,pdfproducer=medialab}
\usepackage[english]{babel}
\usepackage{catchfile}
\usepackage{indentfirst}
\usepackage{graphicx}
\usepackage{enumerate}
\usepackage[yyyymmdd,hhmmss]{datetime}
\newcommand{\email}[1]{\href{mailto:#1}{\tt #1}}

\numberwithin{equation}{section}

\usepackage{catchfile}
\newcommand{\getenv}[2][]{%
  \CatchFileEdef{\temp}{"|kpsewhich --var-value #2"}{}%
  \if\relax\detokenize{#1}\relax\temp\else\let#1\temp\fi}
\getenv[\USER]{USER}

\newcommand{\blue}[1]{\color{blue} #1 \color{black}}

\newcommand{\be}{\begin{equation}}
\newcommand{\ee}{\end{equation}}
\newcommand{\ba} {\begin{equation}\begin{aligned}}
\newcommand{\ea} {\end{aligned}\end{equation}}
\newcommand{\bea}{\begin{eqnarray}}
\newcommand{\eea}{\end{eqnarray}}

\def\Tr{{\rm Tr}}
\def\hc{\mathrm{h.c.}}

\newcommand{\derp}{\partial}

\def\ab{\text{ ab}}
\def\TeV{\text{ TeV}}
\def\GeV{\text{ GeV}}

\newcommand{\LL}{\mathscr{L}}

\newcommand{\cL}{\mathcal{L}}
\newcommand\cO{\mathcal{O}}
\newcommand\cG{\mathcal{G}}

\newcommand\cP{\mathcal{P}}
\newcommand\cT{\mathcal{T}}

\newcommand{\U}{\mathbf{U}}
\newcommand{\V}{\mathbf{V}}
\newcommand{\T}{\mathbf{T}}
\newcommand{\D}{\mathbf{D}}
\newcommand{\F}{\mathcal{F}}

\begin{document}
\renewcommand*{\thefootnote}{\fnsymbol{footnote}}
\begin{titlepage}
\vspace*{-1cm}
{FTUAM-19-9}
\hfill{IFT-UAM/CSIC-19-59}
\hfill{VBSCAN-PUB-03-19}\\
\vskip 2cm
\begin{center}
\mathversion{bold}
\blue{{\LARGE\bf Same-sign $WW$ Scattering in the HEFT:\\[2mm] 
Discoverability vs. EFT Validity}}\\[4mm]
\mathversion{normal}
\vskip .3cm
\end{center}
\vskip 0.5  cm
\begin{center}
{\large\bf P.~Koz\'ow}~$^{a),b)}$\footnote{\email{Pawel.Kozow@fuw.edu.pl}},
{\large\bf L.~Merlo}~$^{c)}$\footnote{\email{luca.merlo@uam.es}},
{\large\bf S.~Pokorski}~$^{a)}$\footnote{\email{Stefan.Pokorski@fuw.edu.pl}},
{\large\bf M.~Szleper}~$^{d)}$\footnote{\email{Michal.Szleper@ncbj.gov.pl}}
\vskip .7cm
{\footnotesize
$^{a)}$~Institute of Theoretical Physics, Faculty of Physics, University of Warsaw,\\
ul. Pasteura 5, PL-02-093 Warsaw, Poland\\
\vskip .1cm
$^{b)}$~CAFPE and Departamento de F\'isica Te\'orica y del Cosmos, Universidad de Granada, Campus de Fuentenueva, E-18071, Granada, Spain\\
\vskip .1cm
$^{c)}$~Departamento de F\'isica Te\'orica and Instituto de F\'isica Te\'orica, IFT-UAM/CSIC,\\
Universidad Aut\'onoma de Madrid, Cantoblanco, 28049, Madrid, Spain\\
\vskip .1cm
$^{d)}$~National Center for Nuclear Research, High Energy Physics Department,\\
ul. Pasteura 7, PL-02-093, Warsaw, Poland\\
}
\end{center}
\vskip 2cm
\begin{abstract}
\justify
Vector boson scatterings are fundamental processes to shed light on the nature of the electroweak symmetry breaking mechanism. Deviations from the Standard Model predictions on the corresponding observables can be interpreted in terms of effective field theories, that however undergo consistency conditions. In this paper, the same-sign $WW$ scattering is considered within the HEFT context and the correct usage of the effective field theory approach is discussed. Regions of the parameters space are identified where a signal of new physics could be measured at HL-LHC with a significance of more than $5\sigma$ and the effective field theory description is consistently adopted. These results are then translated into bounds on the $\xi$ parameter in the composite Higgs scenario.  The discussion on the agreement with previous literature and the comparison with the equivalent analysis in the SMEFT case are also included.
\end{abstract}
\end{titlepage}
\setcounter{footnote}{0}


\renewcommand*{\thefootnote}{\arabic{footnote}}
\section{Introduction}

The LHC discovery of a scalar boson~\cite{Aad:2012tfa,Chatrchyan:2012xdj} compatible with the Higgs particle as predicted in the Standard Model (SM) represents a milestone in the elementary particle physics. Indeed, it may be the first discovered elementary particle with zero spin, if no substructure is found. Experimental collaborations are nowadays strongly involved in studying its properties to confirm the SM predictions or to point out discrepancies due to Beyond SM (BSM) dynamics in the Electroweak (EW) sector. The necessity of New Physics has already been established, i.e. by non-vanishing active neutrino masses or by the existence of the Dark Matter. So, it should not be surprising  if the Electroweak Symmetry Breaking (EWSB) mechanism would finally be shown to differ from what is described by the SM. 

Two main directions of thoughts on the EWSB mechanism have emerged in the last decades: according to the first one, the EWSB mechanism is due to the linearly realised dynamics of the Higgs sector such as in the SM; the alternative consists in a non-linearly realised dynamics of the Higgs sector, typically occurring in Composite Higgs (CH) models. As direct searches at the LHC have so far not observed any new resonance, that are predicted in CH scenarios,  indirect searches  are important  to disentangle these two possibilities. 

A fundamental aspect of indirect searches is the use of an effective description to account for NP effects at low energies. Two different effective field theories for the SM physical degrees of freedom have been constructed, to describe the linear and the non-linear realisation of the Higgs sector dynamics:
\begin{itemize} 
\item[-]
In the so-called SM Effective Field Theory (SMEFT)~\cite{Buchmuller:1985jz,Grzadkowski:2010es}  the $SU(2)_L\times SU(2)_R$ symmetry in the Higgs sector is realised linearly,  with the Higgs field being a $SU(2)_L\times SU(2)_R$ bi-doublet. 
The perturbative expansion is characterised by inverse powers of the cut-off scale $\Lambda$, representing the NP energy scale. Considering only Baryon and Lepton number preserving operators, the SMEFT Lagrangian is written as the sum of different terms,
\be
\LL_\text{SMEFT}=\LL_\text{SM}+\sum_i c^{(6)}_i \cO^{(6)}_i+\sum_i c^{(8)}_i \cO^{(8)}_i+\ldots\,,
\label{eq:SMEFT}
\ee
where $\LL_\text{SM}$ stands for the SM Lagrangian, $c^{(n)}_i$ are dimensionless coefficients and $\cO^{(n)}$ are operators of canonical dimension $n>4$ that encode the suppression $\Lambda^{4-n}$. Dots represent higher order operators in the expansion in inverse powers of the cut-off. 

The SMEFT is renormalisable order by order in the expansion in $\Lambda$: independently from the number of loops of a given diagram,  the quantum corrections calculated from  the Lagrangian truncated at some order of $\Lambda$  and calculated to the same order in $\Lambda$  can be renormalised by the couplings present in the effective Lagrangian. 

Considering observables whose typical energy is much smaller than the cut-off, the higher the canonical dimension of an operator the smaller is its contribution to those observables. Once the typical energy involved is closer to the cut-off, the ordering of the operators in canonical dimensions is not so meaningful anymore and operators of higher dimensions may dominate. In this case, the effective description breaks down and cannot be considered reliable anymore.
\item[-] 
In the  so-called Higgs Effective Field Theory  (HEFT)~\cite{Feruglio:1992wf,Grinstein:2007iv,Contino:2010mh,Alonso:2012px,Alonso:2012pz,Buchalla:2013rka,Brivio:2013pma,Brivio:2014pfa,Gavela:2014vra,Gavela:2014uta,Eboli:2016kko,Brivio:2016fzo,deFlorian:2016spz,Merlo:2016prs,Buchalla:2017jlu,Alonso:2017tdy} the $SU(2)_L\times SU(2)_R$ symmetry is  realized non-linearly, on the three Goldstone bosons eaten up by the gauge fields.
The physical Higgs field may be either a singlet of the diagonal, the so-called custodial, symmetry $SU(2)_C$ (if in an UV completion  the $SU(2)_L\times SU(2)_R$ symmetry is realised linearly at a certain level) or an additional field never forming the Higgs doublet with the three Goldstone bosons (GBs). This Lagrangian is the most general description of the EW and Higgs couplings, satisfying the gauge symmetry of the SM: in specific limits, it may reduce to the SM Lagrangian, or may coincide with the SMEFT one, or may match the description of CH models~\cite{Kaplan:1983fs,Kaplan:1983sm,Banks:1984gj,Agashe:2004rs,Gripaios:2009pe,
Alonso:2014wta,Hierro:2015nna,Feruglio:2016zvt,Gavela:2016vte,Merlo:2017sun,Alonso-Gonzalez:2018vpc} or even dilaton models~\cite{Halyo:1991pc,Goldberger:2008zz,Hernandez-Leon:2017kea}. 

The HEFT can be considered as the merging between the chiral perturbation theory applied to the longitudinal components of the gauge bosons and the Higgs and the SMEFT once dealing with the other SM particles. Following the traditional notation of the EW Chiral Lagrangian (EW$\chi$L)~\cite{Appelquist:1980vg,Longhitano:1980iz,Longhitano:1980tm,Feruglio:1992wf}, the SM GBs $\pi^a$ are described by means of a dimensionless unitary matrix transforming as a bi-doublet of the global $SU(2)_L\times SU(2)_R$ symmetry,
\be
\U(x)=e^{i\,\sigma_a\pi^a(x)/v}\,,\qquad\qquad\U(x)\to L\U(x)R^\dag\,,
\ee
with $v=246\GeV$ being the EW scale defined through the $W$ mass, and $L$ ($R$) denote the $SU(2)_{L(R)}$ transformation. It is customary to introduce two objects, the vector and the scalar chiral fields, that transform in the adjoint of $SU(2)_L$:
\be
\V_\mu(x)\equiv\left(\D_\mu\U(x)\right)\U(x)^\dag\,,\qquad\qquad
\T(x)\equiv\U(x)\sigma_3\U(x)^\dag\,,
\ee
where the covariant derivative reads
\be
\D_\mu\U(x)\equiv\derp_\mu\U(x)+ig\,W_\mu(x)\U(x)-\dfrac{ig'}{2}B_\mu(x)\U(x)\sigma_3\,. 
\ee
While $\V_\mu(x)$ is $SU(2)_C$ conserving, $\T(x)$ is not and therefore plays the role of a custodial symmetry breaking spurion.

Differently than in  the SMEFT, the physical Higgs field does not have to belong to a doublet representation of $SU(2)_L$ and  generic functions $\F(h)$ are used as building blocks to construct the effective operators. These functions are made dimensionless by implicitly weighting the insertions of the Higgs field with an appropriate power of the EW scale $v$.\footnote{The scale associated to the $h$ is typically denoted as $f\neq v$, as in CH models. However, in the HEFT there is the freedom to redefine this scale and, considering that the Higgs physics does not intervene in the present analysis, the EW scale will be taken as the reference one in the functions $\F(h)$ without any loss of generality.} 
In concrete CH models, where the physical Higgs field arises as a pseudo-GB, and in fact is one component of the Higgs doublet,  the functions $\F(h)$ take trigonometric expressions~\cite{Alonso:2014wta,Hierro:2015nna}.

Due to the presence of the vector and scalar chiral fields $\V_\mu$ and $\T$ (the dependence of the space-time coordinates will be omitted in what follows), a single-parameter expansion is meaningless in the HEFT~\cite{Gavela:2016bzc}: the perturbative expansion in the EW$\chi$L is typically based on a derivative counting rule; while, as said above, the SMEFT is ruled by an expansion in inverse powers of the cut-off scale. 

Contrary to the SMEFT, the renormalisability is more involved: in the pure EW$\chi$L, 1-loop diagrams with one insertions of a two-derivative coupling, usually listed in the Leading Order (LO) Lagrangian, produce divergences that require the introduction of operators with four-derivatives, which generically constitute the Next-to-the-Leading Order (NLO) Lagrangian. This is not compatible with the SMEFT renormalisability order by order in $\Lambda$. 

For both SMEFT and HEFT, the distinction in LO, NLO, etc\ldots, sometimes fails in ordering the impact of the different operators. The latter depends on the structure of the operators and on the energy involved in the observables under consideration. Once the energy is smaller but close to the cut-off, a counting based on the so-called primary dimension $d_p$~\cite{Gavela:2016bzc} is useful: it counts the canonical dimension of the leading terms in the expansion of a given object. Indeed, the matrix $\U$ and the functions $\F(h)$ hide the dependence on the scale $v$:
\be
\U=1+2i\dfrac{\sigma_a\pi^a}{v}+\ldots\,,\qquad\qquad
\F(h)=1+2a\dfrac{h}{v}+\ldots\,,
\label{GenericUF}
\ee 
and as a consequence $\U$ and $\T$ has $d_p=0$, while $\V_\mu$ and $\derp_\mu\F(h)$ have $d_p=2$. 

Any operator may be ordered in terms of its $d_p$ and it allows to link the particular structure of an operator to the strength of a physical signal measured by  cross sections. An interesting application is that operators with the same $d_p$ are expected to have similar impact on a given observables: this information may be used to identify the complete set of operators describing in a similar way the same process, although they belong to different orders in the expansion. Another application of the primary dimension is that if the $d_p$ of an HEFT operator is smaller than the canonical dimension of the SMEFT operator that contributes to  a same observable (this SMEFT operator will be refereed to as ``linear sibling'' of this HEFT operator), then the process described by these operators is expected to have a higher cross section in the HEFT than in the SMEFT: this process may be used to test the linearity of the Higgs sector dynamics ~\cite{Brivio:2013pma,Brivio:2014pfa,Gavela:2014vra,Brivio:2015kia,Brivio:2016fzo,Brivio:2017ije}. 
\end{itemize}

The precise determination of both the Higgs couplings and of triple and quartic gauge couplings are fundamental for understanding the dynamics of the EWSB mechanism. Studies of the longitudinal EW vector boson scattering (VBS) may shed some light on the intimate nature of the EW sector and the use of EFTs, in that specific case of the EW$\chi$L, dates back to the late '80s~\cite{Dobado:1989ax,Dobado:1989ue,Dobado:1989gr,Dobado:1990jy,Dobado:1995qy,Dobado:1999xb,Alboteanu:2008my}. After the discovery of a scalar resonance at the LHC, VBS received a renewed attention both from the experimental collaborations~\cite{Aad:2014zda,CMS:2014uib,Khachatryan:2014sta,Aaboud:2016ffv,Aad:2016ett,Sirunyan:2017fvv,Sirunyan:2017ret} and from theorists who analysed possible signals of NP in these processes by means of the SMEFT Lagrangian~\cite{Kilian:2014zja,Kalinowski:2018oxd,Brass:2018hfw,Gomez-Ambrosio:2018pnl}, or of the HEFT one~\cite{Espriu:2012ih,Espriu:2013fia,Delgado:2013loa,Delgado:2013hxa,Espriu:2014jya,Delgado:2014jda,Delgado:2017cls}, or even in a generic CH scenario~\cite{Ballestrero:2009vw,BuarqueFranzosi:2017prc}.

In particular, in Ref.~\cite{Kalinowski:2018oxd} a strategy for same-sign WW scattering at the LHC data analysis was developed in the context of SMEFT, assuming that there will be evidence for New Physics (NP) in the future data (i.e. discrepancies at $\geq5$ standard deviations $5\sigma$ with respect to the SM predictions). Particular emphasis was put on the proper use of EFT in its region of validity. For a detailed discussion see Ref.~\cite{Kalinowski:2018oxd}, while only few details will be given here. 

First of all, certain effective operators cause the $W^+W^+\rightarrow W^+W^+$ scattering amplitude to grow with energy, leading to unitarity violation above a certain WW center of mass (c.o.m.) energy scale, labelled $M_{WW}$ matching the notation in Ref.~\cite{Kalinowski:2018oxd}. The value at which the unitarity is violated, denoted as $M^U$, can be determined in terms of the ratios between the operator coefficients and the cut-off scale, $f_i\equiv c_i/\Lambda^n$, associated to each operator contributing to WW scattering. This value for $M^U$ is then identified with the bound arising from the unitarity condition for the $J=0$ in $WW\rightarrow WW$ on-shell scattering partial wave criterium: $|a_{J=0}<1/2|$. In the region where the c.o.m. energy of the WW system is larger than the unitarity violation scale, $M_{WW}>M^U(f_i)$, the amplitude of the full considered reaction $pp\rightarrow jj\ell\ell'\nu_\ell\nu_{\ell'}$ cannot be consistently described within an EFT approach. This condition also leads to a bound on the cut-off of the EFT: $\Lambda$ must be smaller than the scale at which the unitarity gets violated, i.e. $\Lambda < M^U(f_i)$, as for larger energies new degrees of freedom (d.o.f.) are expected to be added into the spectrum. Moreover, another constraint on $\Lambda$ follows by requiring the validity of the Lagrangian expansion in terms of energy over the cut-off: $\Lambda$ must be larger than the energy involved in the process, that is $\Lambda>M_{WW}$. All in all, the following inequality arises from the conditions for a consistent and valid EFT description of the $WW$ scattering:
\be
M_{WW}<\Lambda<M^U(f_i)\,.
\label{eq:pk}
\ee
For fixed $f_i$, different (continuous) choices of the $\Lambda$ scale can be considered.

A second relevant aspect that was considered in Ref.~\cite{Kalinowski:2018oxd}  is the fact that $M_{WW}$ is not an observable in the 4 lepton WW decay channel,  and various kinematic distributions need be used to describe data. In particular, the kinematic window spans values up to $M_{WW}^\text{max}$, that may be larger than the cut-off $\Lambda$. However, the only relevant data in an EFT fit must belong to the region where $M_{WW}<\Lambda$: only in this case, physical conclusions can be drawn from the analysis; in particular the tail of the $M_{WW}$ distribution should not have a significant impact on the fit. This condition, that has not been considered in previous similar analyses, has been implemented quantitatively in the analysis in Ref.~\cite{Kalinowski:2018oxd} introducing two signal estimates. The first, labelled $D_i^\text{EFT}$, defines signals coming uniquely from the EFT in its range of validity and assumes only the SM contribution in the region where $\Lambda<M_{WW}<M_{WW}^\text{max}$: it reads
\begin{equation}
D_i^\text{EFT}=\int^{\Lambda}_{2M_W}\frac{d\sigma}{dM}\Big|_\text{model} dM +\int_{\Lambda}^{M^\text{max}_{WW}}\frac{d\sigma}{dM}\Big|_\text{SM} dM\,,
\label{dsigma}
\end{equation}
where the index $i$ refers to a specific choice for $f_i$. In the second estimate, a tail regularisation of the distribution in $M_{WW}$ is considered, and the choice in Ref.~\cite{Kalinowski:2018oxd} was of a constant $W^+W^+\rightarrow W^+W^+$ amplitude above $\Lambda$. In particular, this method guarantees that the partial wave unitarity condition is not violated above $\Lambda$. This second estimate reads:
\begin{equation}
D_i^\text{BSM}=\int^{\Lambda}_{2M_W}\frac{d\sigma}{dM}\Big|_\text{model} dM +\int_{\Lambda}^{M^\text{max}_{WW}}\frac{d\sigma}{dM}\Big|_{A= \text{const}} dM
\label{unitarized}
\end{equation}
where $A= \text{const}$ denotes the regularisation. In order to guarantee that also with this second signal estimate the physical effects in the fit do not come from the tail of the distribution, statistical consistency between the two estimates within two standard deviations ($2\sigma$) was required. 

Summarising, in the analysis for the SMEFT Lagrangian presented in Ref.~\cite{Kalinowski:2018oxd}, the $(f_i,\Lambda)$ parameter space was studied requiring i) $\geq 5\sigma$ discrepancy between $D^{BSM}_i$ and the SM prediction, ii) the condition in Eq.~\eqref{eq:pk}, and that iii) the distributions $D^\text{EFT}_i$ and $D^\text{BSM}_i$ do not differ more than $2\sigma$ from each other. The area selected by these three requirements corresponds to the discovery region of a certain class of EFT that contributes to the WW scattering process: the larger the region the larger the chance for sensible description of the forthcoming data within the EFT approach. In Ref.~\cite{Kalinowski:2018oxd}, a single operator analysis has been performed, that is considering the SM Lagrangian with the addition of a single $d=8$ SMEFT operator at a time that modifies the WWWW interaction: the discovery regions have been determined for each $d=8$ SMEFT operator included in the Lagrangian. Typically, the form of the discovery regions is similar to a triangle in the $(f_i,\Lambda)$ plane, delimited by the unitarity limit from above, the $5\sigma$ BSM observability from the left and the $2\sigma$ statistical consistency between the two signal estimates from the right.

The aim of the present paper is to apply the analysis performed in Ref.~\cite{Kalinowski:2018oxd} to the HEFT description, determining the discovery potential of several operators that modify the WWWW interaction considering HL-LHC ($14\TeV$ $pp$ collision energy and $3 \ab^{-1}$ of integrated luminosity). As for the SMEFT, only a single operator analysis will be performed, considering a specific set of $d_p=8$ HEFT operators that {\it genuinely} affect the WW scattering, that is do not contribute to triple gauge boson interactions. 

As already mentioned, the effects of HEFT operators to VBS have already been studied in the literature~\cite{Espriu:2012ih,Espriu:2013fia,Delgado:2013loa,Delgado:2013hxa,Espriu:2014jya,Delgado:2014jda,Delgado:2017cls}, but without considering explicitly the conditions listed above, that is $5\sigma$ discoverability and the consistency of the EFT approach once enforcing that the distribution tail does not have relevant (at $2\sigma$) effects on the signal.

The results obtained adopting the HEFT Lagrangian can then be interpreted in terms of specific Ultraviolet (UV) completions. In particular, the focus will be on the CH context, providing a bound on the characteristic parameter $\xi\equiv v/f$, being $f$ the scale of the global symmetry breaking. This bound is complementary to those obtained from EW precision observables or from the non-discovery of composite resonances (for a recent review see Ref.~\cite{Panico:2015jxa}). 

The rest of the paper is organised as follows. In Sect.~\ref{Sect.HEFTOps}, the operators of the HEFT relevant to discuss the VBS will be introduced. The results of the analysis will be presented and discussed in Sect.~\ref{Sect.Results}. Final remarks can be found in Sect.~\ref{Sect.Conclusions}, while more technical details are left for the appendix~\ref{App.A}.

\boldmath
\section{The HEFT Operators Relevant for VBS}
\label{Sect.HEFTOps}
\unboldmath

The HEFT Lagrangian can be written as the sum of two terms,
\be
\LL_\text{HEFT}\equiv\LL_0+\Delta\LL\,,
\ee
where the first term contains the LO operators and the second one describes new interactions and deviations from the LO contributions. There is no common agreement on which operators belongs to $\LL_0$ and which to $\Delta\LL$: even in the choice of $\LL_0$ there are different opinions in the literature. Following Ref.~\cite{Brivio:2016fzo} and the recipe illustrated in the Introduction, the LO Lagrangian contains the kinetic terms for all the particles in the spectrum, the Yukawa couplings and the scalar potential: restricting to CP conserving couplings,
\be
\begin{split}
\LL_0=&
-\dfrac{1}{4}\cG_{\mu\nu}^\alpha\cG^{\alpha\,\mu\nu}
-\dfrac{1}{4}W_{\mu\nu}^aW^{a\,\mu\nu}
-\dfrac{1}{4}B_{\mu\nu}B^{\mu\nu}+\\
&
+\dfrac{1}{2}\derp_\mu h\derp^\mu h
-\dfrac{v^2}{4}\Tr\left(\V_\mu\V^\mu\right)\F_C(h)
-V(h)+\text{\tt fermions}\,,
\end{split}
\label{L0}
\ee
where $\tt fermions$ refers to all the terms involving fermions and that will not be considered here as they do not enter the present analysis.
The first line describes the kinetic terms for the gauge bosons, with the color (weak) index with Greek (Latin) letters. The second line contains the Higgs and the GBs kinetic term, the scalar potential, and the mass terms for $W$ and $Z$ gauge bosons. 

The function $\F_C(h)$ in the second line of Eq.~(\ref{L0}) is conventionally written as
\be
\F_C(h) = 1 + 2 a_C \frac{h}{v}+b_C \frac{h^2}{v^2}+\dots\,,
\label{FC}
\ee
where the dots refer to higher powers in $h/v$. In the SM case, the first two coefficients of $\F_C(h)$ are exactly equal to $a_C=1=b_C$, while the ones corresponding to higher orders are identically vanishing. On the experimental side, present fits indicate a central value of these coefficients close to $0.9$, compatible with the SM within the $1\sigma$ uncertainty~\cite{Brivio:2016fzo}. Although in a general analysis $a_C$ and $b_C$ are free parameters, the SM values will be assumed in this analysis, being the focus on the effects of genuine quartic operators into $WWWW$ scatterings.

The second part of the HEFT Lagrangian, $\Delta\LL$ contains all the color and EW invariant operators appearing beyond the LO, including corrections to $\LL_0$ and new couplings. In what follows, only a set of operators will be considered, that are useful to discuss the dominant contributions to $WW\to WW$ process: adopting the notation of Refs.~\cite{Brivio:2016fzo,Eboli:2016kko}; (in particular $\Lambda$is the EFT cut-off scale),
\be
\Delta\LL\supset \sum_{i=6,11}c_i\cP_i\F_i(h)+\sum_{i=42,43,44,61,62}c_{i}\cT_i\F_i(h)+\sum_{i=0,1,2}c_{Ti}\cO_{Ti}\F_i(h)\,,
\label{DeltaLHEFT}
\ee
where $c_i$ are free coefficients and the operators are defined as
\be
\begin{aligned}
\cP_{6}=&\dfrac{1}{16\pi^2} \Tr(\V_\mu\V^\mu)\Tr(\V_\nu\V^\nu)\\
\cP_{11}=&\dfrac{1}{16\pi^2}  \Tr(\V_\mu\V_\nu)\Tr(\V^\mu\V^\nu) \\[5mm]
\cT_{42}=&\dfrac{1}{\Lambda^2}\Tr(\V_\alpha W_{\mu\nu})\Tr(\V^\alpha W^{\mu\nu})\\
\cT_{43}=&\dfrac{1}{\Lambda^2}\Tr(\V_\alpha W_{\mu\nu})\Tr(\V^\nu W^{\mu\alpha})\\
\cT_{44}=&\dfrac{1}{\Lambda^2}\Tr(\V^\nu W_{\mu\nu})\Tr(\V_\alpha W^{\mu\alpha})\\
\cT_{61}=&\dfrac{1}{\Lambda^2}W^a_{\mu\nu}W^{a\mu\nu}\Tr(\V_\alpha\V^\alpha )\\
\cT_{62}=&\dfrac{1}{\Lambda^2}W^a_{\mu\nu}W^{a\mu\alpha}\Tr(\V_\alpha\V^\nu )\\[5mm]
\cO_{T_0}=&\dfrac{16\pi^2}{\Lambda^4}W^a_{\mu\nu}W^{a\mu\nu}W^b_{\alpha\beta}W^{b\alpha\beta}\\
\cO_{T_1}=&\dfrac{16\pi^2}{\Lambda^4}W^a_{\alpha\nu}W^{a\mu\beta}W^b_{\mu\beta}W^{b\alpha\nu}\\
\cO_{T_2}=&\dfrac{16\pi^2}{\Lambda^4}W^a_{\alpha\mu}W^{a\mu\beta}W^b_{\beta\nu}W^{b\nu\alpha}
\end{aligned}
\label{HEFTOps}
\ee
being $W_{\mu\nu}\equiv W_{\mu\nu}^a\sigma^a/2$. These operators are particularly interesting because they do not contribute to triple gauge vertices, which are strongly constrained: in this case, the analysis that has the best chances to provide interesting constraints on these operators is on VBS. These operators are labelled in the literature as genuine quartic operators. To facilitate the matching between this notation and the one used in previous literature, the operators $\cP_{6}$ and $\cP_{11}$ are also known as $\cL_5$ and $\cL_4$, respectively, of the EW$\chi$L.

The numerical pre-factors in these operators are assigned according to the Naive Dimensional Analysis (NDA) master formula first introduced in Ref.~\cite{Manohar:1983md} and later modified in Refs.~\cite{Brass:2018hfw,Gavela:2016bzc}. Following the notation of Ref.~\cite{Gavela:2016bzc}:
\be
\frac{\Lambda^4}{16 \pi^2 } \left[\frac{\partial}{\Lambda}\right]^{N_p}  
\left[\frac{ 4 \pi\,  \phi}{ \Lambda} \right]^{N_\phi}
 \left[\frac{ 4 \pi\,  A}{ \Lambda } \right]^{N_A}  
\left[\frac{ 4 \pi \,  \psi}{\Lambda^{3/2}}\right]^{N_\psi} 
\left[ \frac{g}{4 \pi }  \right]^{N_g}
\left[\frac{y}{4 \pi } \right]^{N_y}\,,
\label{MasterFormula}
\ee
where $\phi$ represents either the SM GBs or the physical Higgs $h$, $\psi$ a generic fermion, $A$ a generic gauge field, $g$ the generic gauge coupling, $y$ the generic Yukawa coupling, while $N_i$ refer to the number of times each field appears in a given operator. Therefore, the factors associated to $\V_\mu$ and to $W_{\mu\nu}$ are 
\be
\left[\frac{\V_\mu}{\Lambda}\right]^{N_{\V_\mu}}\left[\frac{4\pi}{\Lambda^2}W_{\mu\nu}\right]^{N_{W_{\mu\nu}}}\,.
\ee
Also $\LL_0$ is normalised according to this formula, with the exception of the term proportional to $\F_C(h)$ and the Yukawa terms that present the well-known fine-tuning problem typical of theories with a non-linearly realised EWSB mechanism. The NDA normalisation is very useful because relates the values of the Wilson coefficients to the weak or strong interacting phase of the considered theory: if a Wilson coefficients turns out to be equal of larger than $1$, then the corresponding interactions become strongly coupled; while if it is smaller than $1$, the corrections induced by the corresponding interactions are subdominant, a sign of a weakly coupled theory. 

In papers where these operators are listed according to the number of derivatives, that is the typical counting of the EW$\chi$L, they belong to three different groups: $\cP_6$ and $\cP_{11}$ are listed among the $\cO(p^4)$ operators; $\cT_{42}$, $\cT_{43}$, $\cT_{44}$, $\cT_{61}$ and $\cT_{62}$ are considered $\cO(p^6)$; finally, $\cO_{T_0}$, $\cO_{T_1}$ and $\cO_{T_2}$ are inserted in the $\cO(p^8)$ group. However, at the phenomenological level, the primary dimension is what matters to establish the impact of an operator, as already discussed in the Introduction. All these operators have $d_p=8$ and therefore the whole set should be taken into consideration.

Ref.~\cite{Eboli:2016kko} presents a much longer list of operators which present a similar structure to the ones listed in Eq.~(\ref{HEFTOps}). However, any operator containing the scalar chiral field $\T$ do not contain interaction between four $W$ and therefore should not be considered in this analysis. The other operators that are not listed above and do not  contain $\T$ have higher primary dimension and therefore are expected to provide subdominant contributions.

After decomposing the operators listed in Eq.~(\ref{HEFTOps}) in terms of the Lorentz structures that they describe, it  is useful to check the independence of all these terms once focusing only on the $WW\to WW$ scattering and to compare with the corresponding operators in the SMEFT case. Reporting only the $WWWW$ terms~\cite{Eboli:2016kko},
\be
\begin{aligned}
\cP_{6}\rightarrow&\dfrac{g^4}{16\pi^2}\,W^{+\mu}W^{-}_{\mu}W^{+\nu}W^{-}_{\nu}\\
\cP_{11}\rightarrow&\dfrac{g^4}{32\pi^2}\left(W^{+\mu}W^{-}_{\mu}W^{+\nu}W^{-}_{\nu}+
W^{+\mu}W^{-\nu}W^{+}_{\mu}W^{-}_{\nu}\right)\\[5mm]
\cT_{42}\rightarrow&-\dfrac{g^2}{4\Lambda^2}\left[2W^{+\mu\nu}W^{-}_{\mu\nu}W^{+\alpha}W^{-}_{\alpha}+
\left(W^{+\mu\nu}W^{+}_{\mu\nu}W^{-\alpha}W^{-}_{\alpha}+\hc\right)\right]\\
\cT_{43}\rightarrow&-\dfrac{g^2}{4\Lambda^2}\left(W^{+\mu\nu}W^{-}_{\mu\alpha}W^{+}_\nu W^{-\alpha}+
W^{+\mu\nu}W^{+}_{\mu\alpha}W^{-\alpha}W^{-}_{\nu}+\hc\right)\\
\cT_{44}\rightarrow&-\dfrac{g^2}{4\Lambda^2}\left(W^{+\mu\nu}W^{-}_{\mu\alpha}W^{+\alpha} W^{-}_{\nu}+
W^{+\mu\nu}W^{+}_{\mu\alpha}W^{-\alpha}W^{-}_{\nu}+\hc\right)\\
\cT_{61}\rightarrow&-\dfrac{2g^2}{\Lambda^2}\,W^{+\mu\nu}W^{-}_{\mu\nu}W^{+\alpha}W^{-}_{\alpha}\\
\cT_{62}\rightarrow&-\dfrac{g^2}{2\Lambda^2}\left(W^{+\mu\nu}W^{-}_{\mu\alpha}W^{+}_\nu W^{-\alpha}+
W^{+\mu\nu}W^{-}_{\mu\alpha}W^{+\alpha} W^{-}_{\nu}+\hc\right)\\[5mm]
\cO_{T_0}\rightarrow&\dfrac{64\pi^2}{\Lambda^4}W^{+\mu\nu}W^{-}_{\mu\nu}W^{+\alpha\beta}W^{-}_{\alpha\beta}\\
\cO_{T_1}\rightarrow&\dfrac{64\pi^2}{\Lambda^4}W^{+\alpha\nu}W^{-}_{\mu\beta}W^{+\mu\beta}W^{-}_{\alpha\nu}\\
\cO_{T_2}\rightarrow&\dfrac{64\pi^2}{\Lambda^4}W^{+\alpha\mu}W^{-}_{\mu\beta}W^{+\beta\nu}W^{-}_{\nu\alpha}\,,
\end{aligned}
\label{HEFTOpsLorentz}
\ee
where $W^\pm_{\mu\nu}\equiv\derp_\mu W^\pm_\nu-\derp_\nu W^\pm_{\mu}$. Ten different Lorentz structure appear on the RH side of these expressions and therefore the ten operators listed in Eq.~(\ref{HEFTOps}) are independent. 

\subsection{Comparison with the SMEFT}
\label{sec:comparison}

Following the analysis in Ref.~\cite{Kalinowski:2018oxd}, there are eight independent operators\footnote{The operator $\cO_{M_6}$ appearing in Eq.~(2.5) of Ref.~\cite{Kalinowski:2018oxd} is not independent once focussing on the $WWWW$ observables.} of canonical dimension $d=8$ that contribute genuinely to the VBS. The complete list reads
\be
\begin{aligned}
\cO_{S_0}=&\dfrac{16\pi^2}{\Lambda^4}\left[\left(D_\mu\Phi\right)^\dag D_\nu\Phi\right]\left[\left(D^\mu\Phi\right)^\dag D^\nu\Phi\right]\\
\cO_{S_1}=&\dfrac{16\pi^2}{\Lambda^4}\left[\left(D_\mu\Phi\right)^\dag D^\mu\Phi\right]\left[\left(D_\nu\Phi\right)^\dag D^\nu\Phi\right]\\
\cO_{M_0}=&\dfrac{16\pi^2}{\Lambda^4}W^a_{\mu\nu}W^{a\mu\nu}\left[\left(D_\alpha\Phi\right)^\dag D^\alpha\Phi\right]\\
\cO_{M_1}=&\dfrac{16\pi^2}{\Lambda^4}W^a_{\mu\nu}W^{a\nu\alpha}\left[\left(D_\alpha\Phi\right)^\dag D^\mu\Phi\right]\\
\cO_{M_7}=&\dfrac{16\pi^2}{\Lambda^4}\left(D_\mu\Phi\right)^\dag W_{\alpha\nu}W^{\alpha\mu} D^\nu\Phi
\end{aligned}
\label{SMEFTOps}
\ee
with the addition of $\cO_{T_0}$, $\cO_{T_1}$ and $\cO_{T_2}$, already defined in Eq.~(\ref{HEFTOps}). In the previous expression, $\Phi$ stands for the SM Higgs doublet and its covariant derivative is defined as
\be
D_\mu\Phi=\left(\derp_\mu+\dfrac{i}{2}g'\,B_\mu+\dfrac{i}{2}g\,\sigma^a W_\mu^a\right)\Phi\,.
\ee
In contrast with Ref.~\cite{Kalinowski:2018oxd}, the normalisation chosen here is the NDA one according with Eq.~(\ref{HEFTOps}).

Considering explicitly the $WWWW$ Lorentz structures allows to identify a correlation between the HEFT and SMEFT operators:
\be
\begin{aligned}
c_6\cP_6\qquad\Longleftrightarrow&\qquad c^{(8)}_{S_1}\cO_{S_1}\\
c_{11}\cP_{11}\qquad\Longleftrightarrow&\qquad c^{(8)}_{S_0}\cO_{S_0}+c^{(8)}_{S_1}\cO_{S_1}\\
c_{61}\cT_{61}\qquad\Longleftrightarrow&\qquad c^{(8)}_{M_0}\cO_{M_0}\\
c_{62}\cT_{62}\qquad\Longleftrightarrow&\qquad c^{(8)}_{M_1}\cO_{M_1}\,.
\end{aligned}
\label{eq:correlations}
\ee
The operators $\cO_{T_0}$, $\cO_{T_1}$ and $\cO_{T_2}$ belong to both the bases and therefore the correlation is trivial. The SMEFT operator $\cO_{M_7}$ contains only part of the interactions described by $\cT_{43}$: the other interactions are described by SMEFT with higher dimensions. Finally, the HEFT operators $\cT_{42}$ and $\cT_{44}$ do not have any correspondence with any of the SMEFT of dimension 8.

\boldmath
\section{Analysis and Results}
\label{Sect.Results}
\unboldmath

This section is devoted to the description of the analysis and to the presentation of the results. The main goal is to determine the discovery regions associated to the genuine quartic coupling HEFT operators listed in Eq.~\eqref{HEFTOps}. As described in the Introduction, three conditions determine the discovery regions: i) $\geq 5\sigma$ discrepancy between the signal estimate $D^{BSM}_i$ defined in Eq.~\eqref{unitarized} and the SM prediction, ii) validity of the EFT description embedded in the condition of Eq.~\eqref{eq:pk}, and iii) consistency of the EFT approach implemented by requiring the $2\sigma$ statistical consistency between the two distributions $D^\text{EFT}_i$ and $D^\text{BSM}_i$.

Samples of $6\times 10^5$ events consistent with the VBS topology for the process $pp\rightarrow jj\mu^+\mu^+\nu\nu$ are generated in MadGraph5\_aMC@NLO~\cite{Alwall:2014hca} v2.6.2 at LO at 14 TeV $pp$ collision energy. The HEFT Lagrangian implementation is obtained with a UFO file~\cite{Degrande:2011ua}, created using the FeynRules~\cite{Alloul:2013bka} package v2.3.32. Cross sections at the output of MadGraph are multiplied by a factor 4 to account for all the lepton (electron and/or muon) combinations in the final state. Hadronization is performed with Pythia v8.2, run within MadGraph. Event files at the reconstructed level are generated with the help of the MadAnalysis5 v1.6.33 package (available within MadGraph). Within the latter the FastJet v3.3.0 package is used with the jet clustering \verb+antikt+ algorithm with \verb+radius=0.35+ and \verb+ptmin=20+; the detector efficiencies is set to 100\%. 

The EFT Lagrangian considered in the analysis is defined as the sum of the SM Lagrangian, by fixing $a_C=1=b_C$ in Eq.~\eqref{FC} and neglecting any higher order contribution in powers of $h/v$, plus $\Delta\LL$ introduced in Eq.~\eqref{DeltaLHEFT}. The SM predictions are then recovered selecting $c_i=0$ for all $i$. The analysis including BSM effects is performed considering only one operator of $\Delta\LL$ at a time, i.e. switching off all the other operator coefficients. 

The SM process $pp \to jj\ell^+\ell^+\nu\nu$ is treated as the irreducible background, while "signal" is defined as the enhancement of the event yield relative to the SM prediction in the presence of a given $d_p=8$ HEFT operator . No reducible backgrounds are simulated.

\begin{figure}[h!] 
  \begin{tabular}{cc}
      \includegraphics[width=0.5\linewidth]{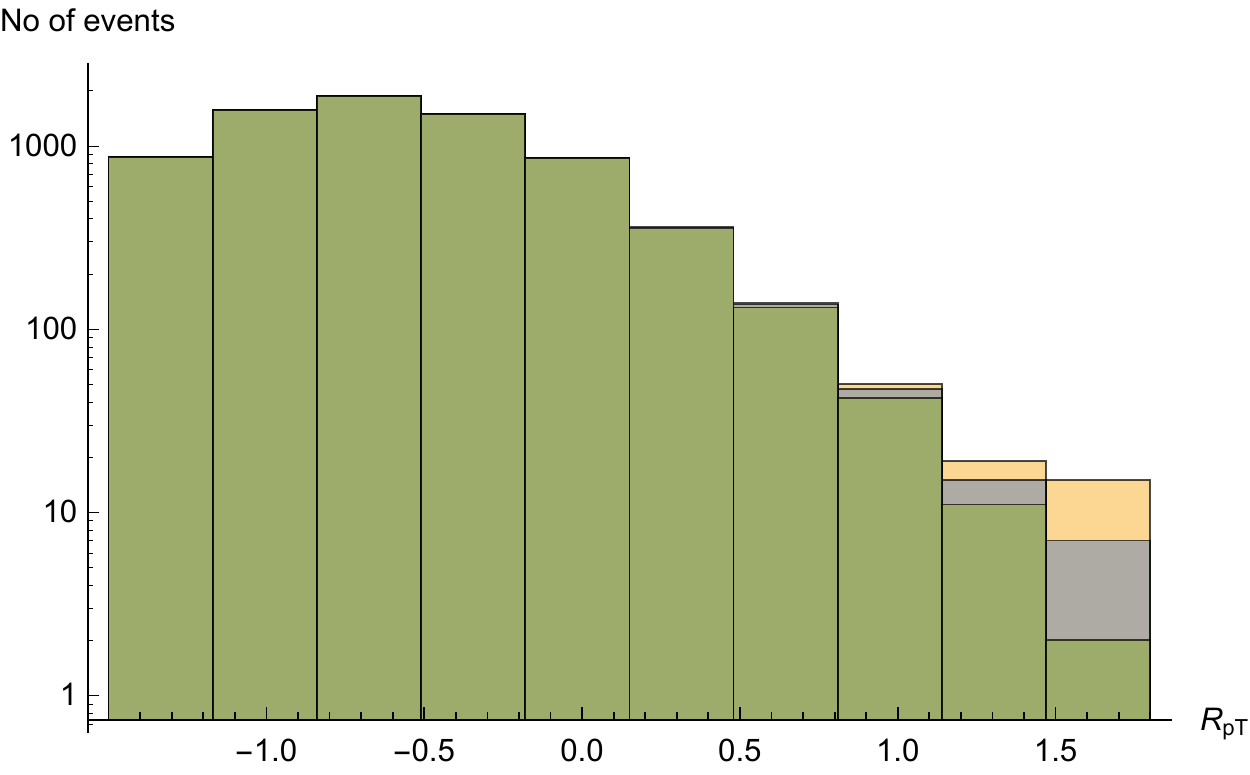} 
& \includegraphics[width=0.5\linewidth]{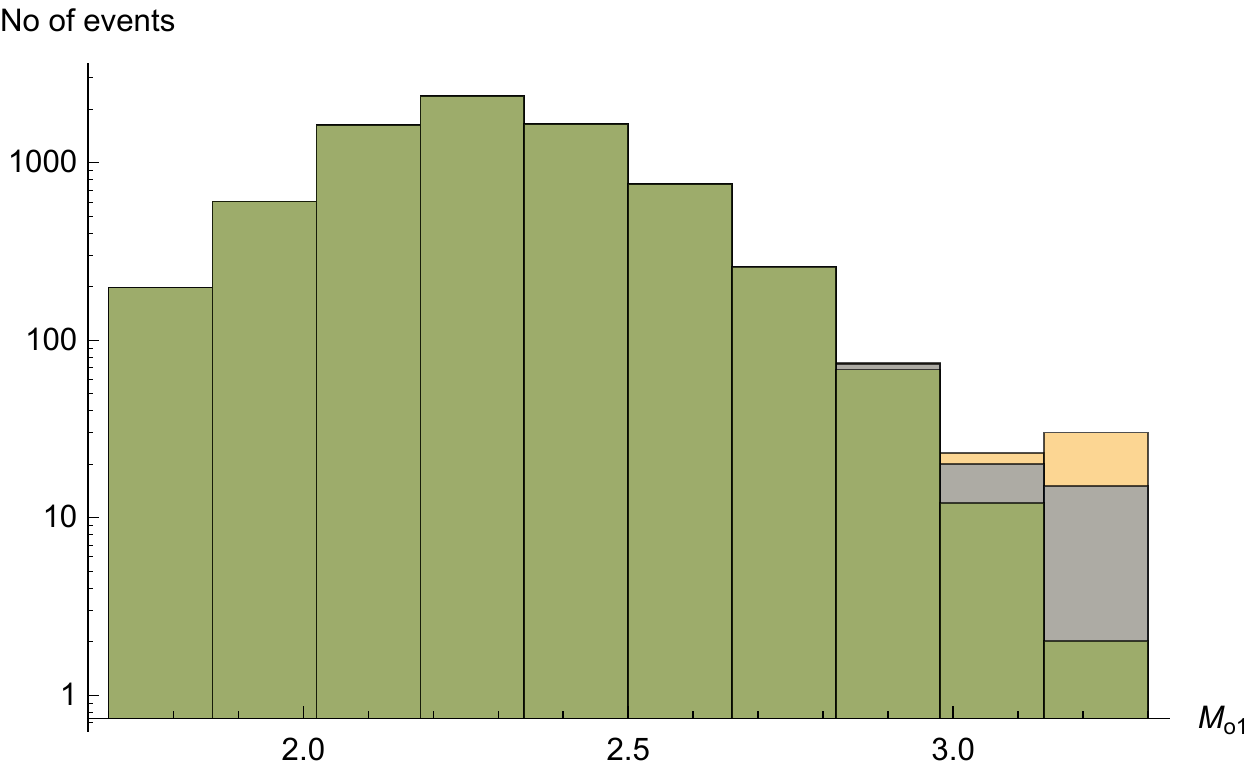} \\
  \end{tabular}
\caption{\em Examples of kinematic distributions used for the assessment of BSM signal significances.
Shown are the distributions $R_{pT}$ and $M_{o1}$ (in log scale): Standard Model is green; the region in yellow corresponds to the operators $\cT_{42}$, taking $f_{42}=0.0075 \TeV^{-2}$ and the high-$M_{WW}$ tail treatment according to Eq.~(\ref{unitarized}); the region in gray corresponds to the operator $\cO_{T1}$, taking $f_{T1}=0.1 \TeV^{-4}$ and the high-$M_{WW}$ tail treatment according to Eq.~(\ref{dsigma}). The value of $\sqrt{s} = 14 \TeV$ and an integrated luminosity of $3 \ab^{-1}$ are assumed.}
\label{fig:kinDist}
\end{figure}

Following Ref.~\cite{Kalinowski:2018oxd}, the event selection criteria consist in requiring at least two reconstructed jets and exactly two leptons (muons or electrons) satisfying the following conditions: $M_{jj} >$ 500 GeV,  $\Delta\eta_{jj} >$ 2.5, $p_T^{~j} >$ 30 GeV, $|\eta_j| <$ 5, $p_T^{\ell} >$25 GeV and $|\eta_\ell| <$ 2.5, being $\eta_{j,\,\ell}$ the pseudorapidity of jets $j$ or leptons $\ell$, respectively. The total BSM signal (Eq.~\eqref{unitarized}) is estimated by suppressing the high-mass tail above the assumed value of $\Lambda$ by applying an additional weight of the form $(\Lambda/M_{WW})^4$ to each generated event in this region,
whereas the EFT-controlled signal (Eq.~\eqref{dsigma}) is calculated by replacing the generated high-mass tail with the one expected in the SM. Signal significances are computed as the square root of a $\chi^2$ resulting from a bin-by-bin comparison of the event yields in the binned distributions of different kinematic observables. Fig.~\ref{fig:kinDist} shows an example of binning, where the most sensitive kinematic variables are considered: 
\be
R_{p_T} \equiv p_T^{\ell1}p_T^{\ell2}/(p_T^{j1}p_T^{j2})\,,
\ee
especially for $\cP_{6}$ and $\cP_{11}$, and 
\be
M_{o1} \equiv \sqrt{(|\vec{p}_T^{~\ell1}|+|\vec{p}_T^{~\ell2}| +|\vec{p}_T^{~miss}|)^2 - (\vec{p}_T^{~\ell1}+\vec{p}_T^{~\ell2}+\vec{p}_T^{~miss})^2}\,,
\ee 
especially for the remaining operators. 

The unitarity limit is determined by applying the T-matrix diagonalised $J=0$ partial wave elastic on-shell WW scattering unitarity criterium, curing the coulomb singularity by a cut of $1\deg$ in the forward and backward scattering regions. The results are cross-checked with VBFLO 1.4.0, founding an agreement within $<5\%$ in the unitarity limits, taken as the lower value between $W^+W^+$ and $W^+W^-$ (deviations in both scatterings are governed by the same Wilson coefficient).

\begin{figure}[h!] 
\begin{tabular}{cc}
\includegraphics[width=0.5\linewidth]{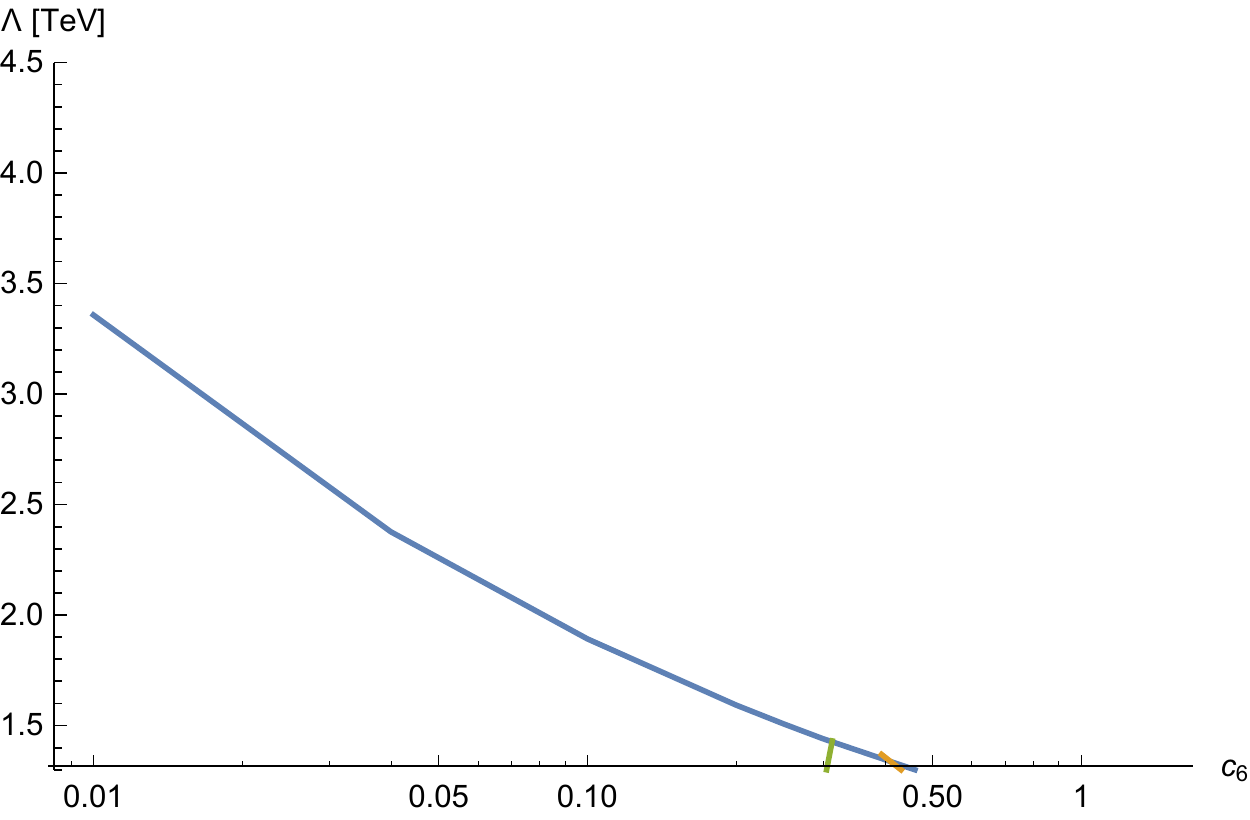} 
& \includegraphics[width=0.5\linewidth]{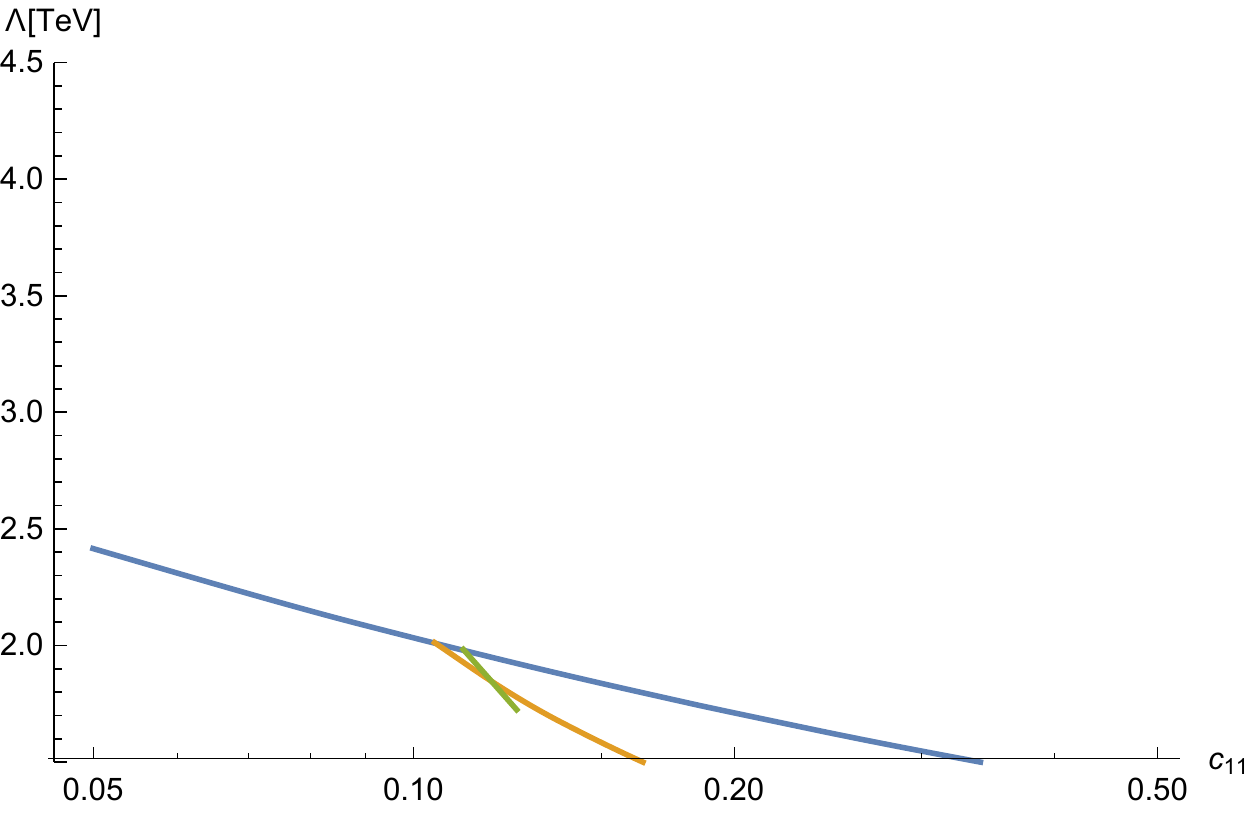} \\
\includegraphics[width=0.5\linewidth]{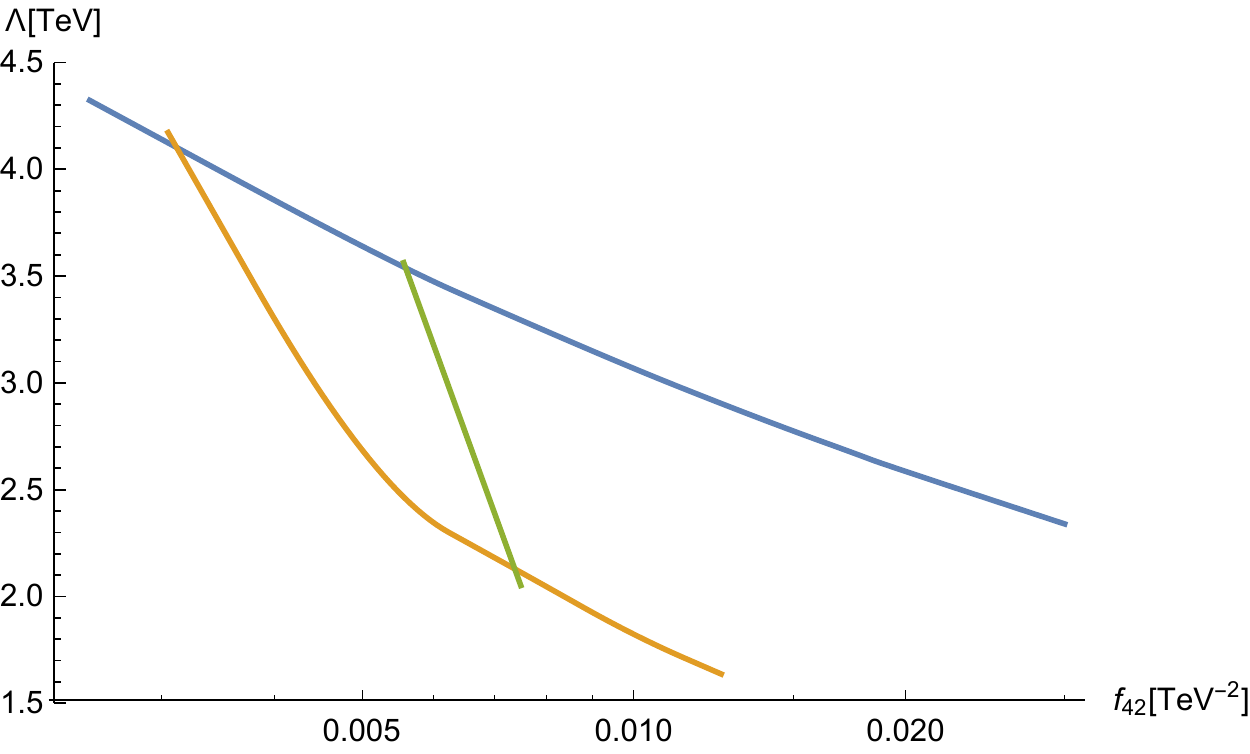} 
& \includegraphics[width=0.5\linewidth]{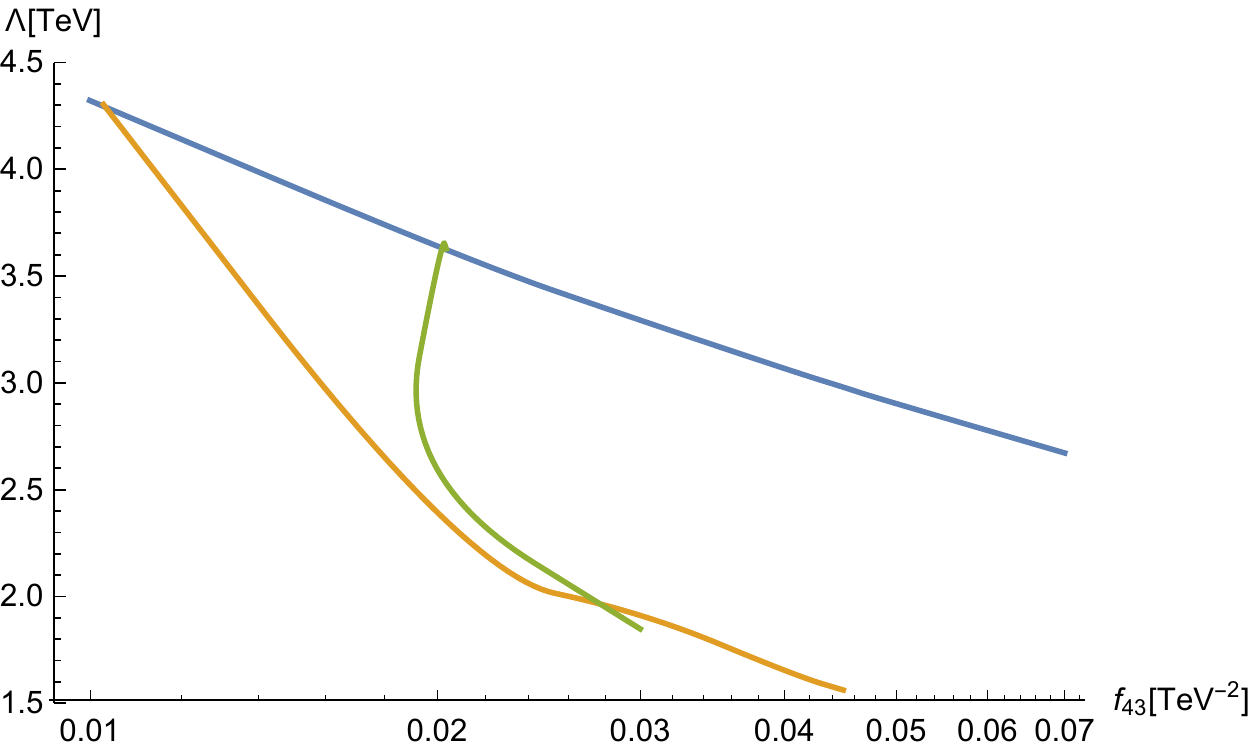} \\
\includegraphics[width=0.5\linewidth]{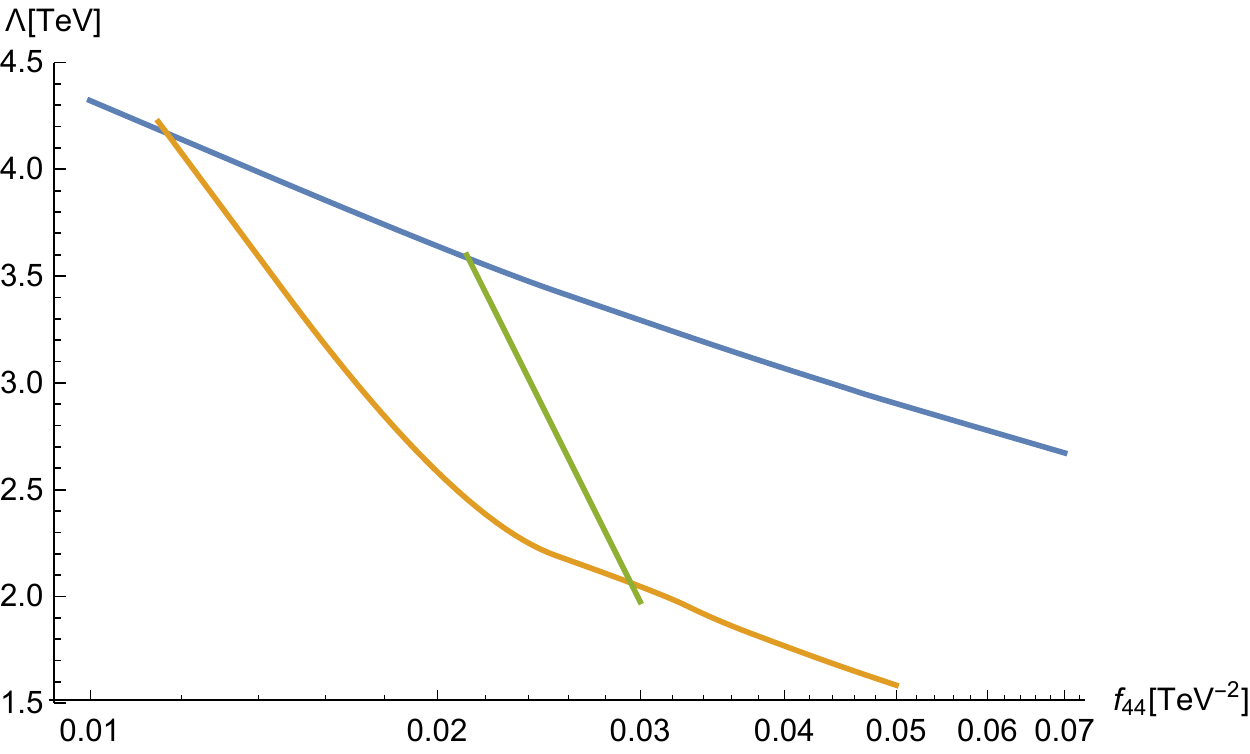} 
& \includegraphics[width=0.5\linewidth]{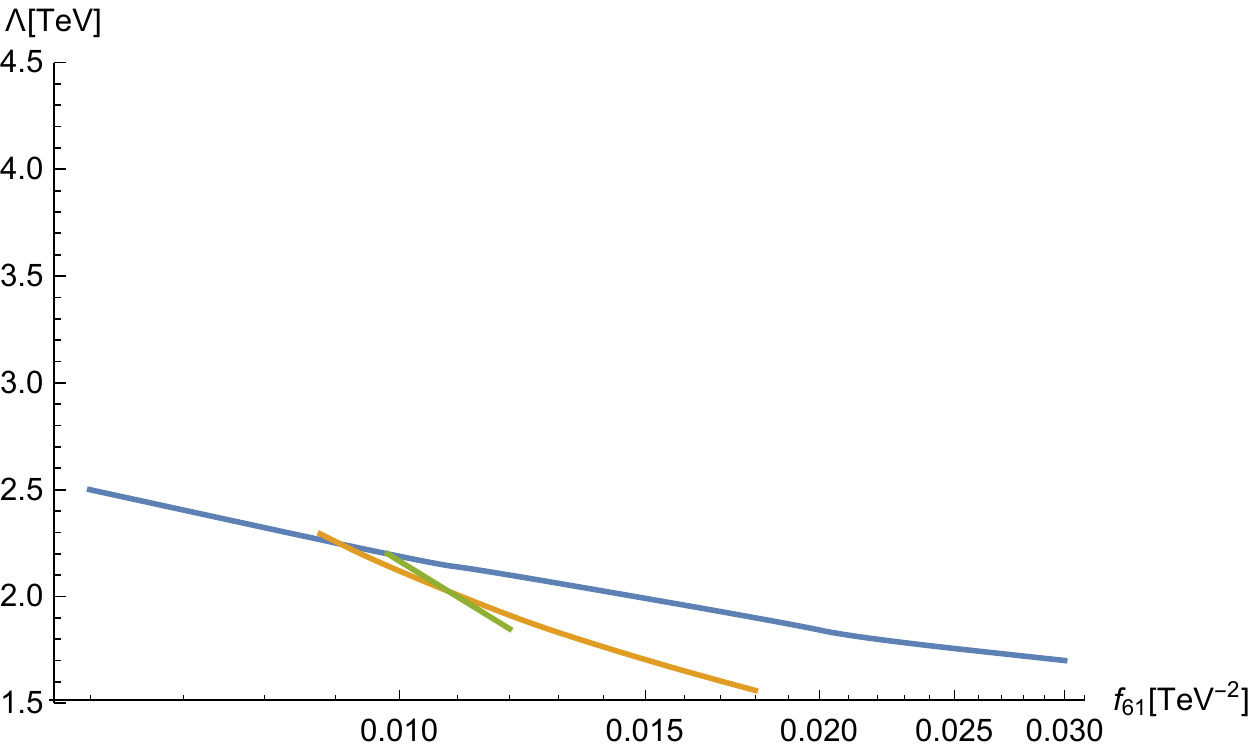} 
\end{tabular}
\caption{\em Regions in the $\Lambda$ {\it vs.} $c_i$ or $f_i$ (with $c_i,\,f_i>0$) space for $d_p=8$ HEFT operators in which a $5\sigma$ BSM signal can be observed and the EFT is applicable. The unitarity limit is shown in blue, the lower limits for a 5$\sigma$ signal significance from Eq.~(\ref{unitarized}) is in yellow, the upper limit on $2\sigma$ EFT consistency in green. $\sqrt{s} = 14 \TeV$ and an integrated luminosity of 3 $ab^{-1}$ are assumed. From top to bottom and from left to right, the operators considered are $\cP_6$, $\cP_{11}$, $\cT_{42}$, $\cT_{43}$, $\cT_{44}$ and $\cT_{61}$.}
\label{fig:trianglesPositivePart1}
\end{figure}

\begin{figure}[h!] 
\begin{tabular}{cc}
\includegraphics[width=0.5\linewidth]{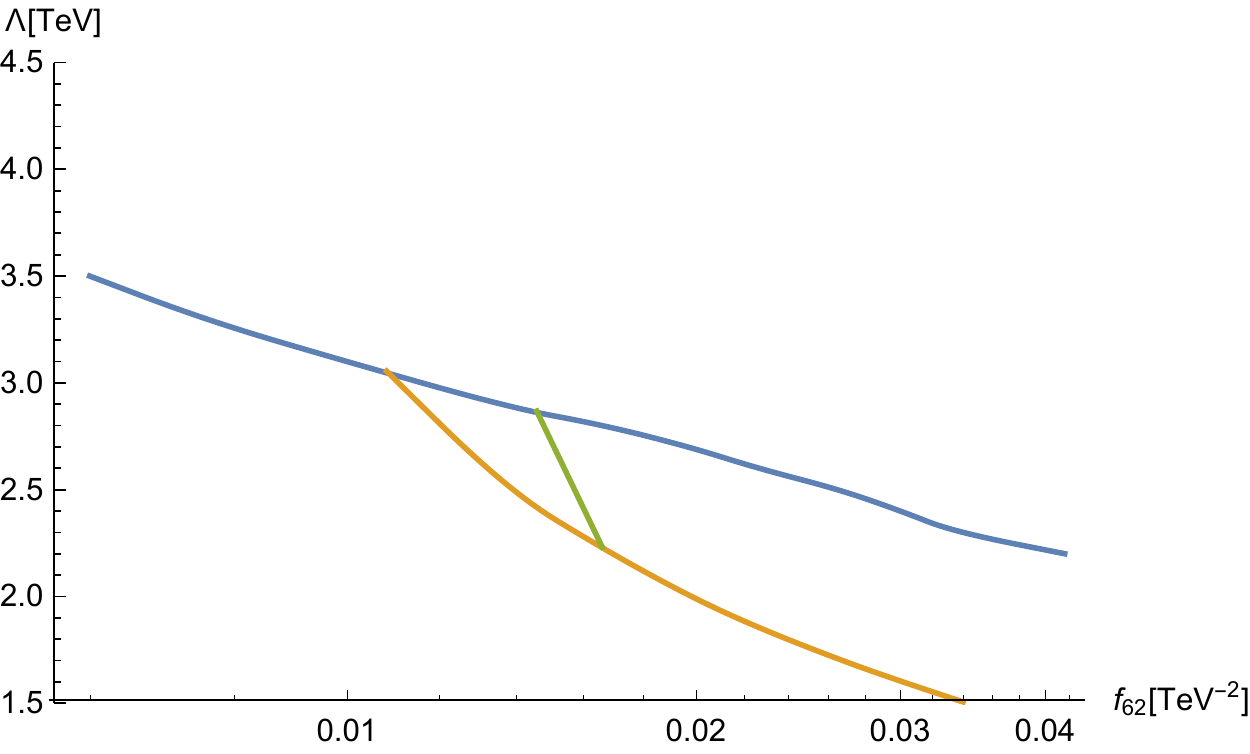}  
&\includegraphics[width=0.5\linewidth]{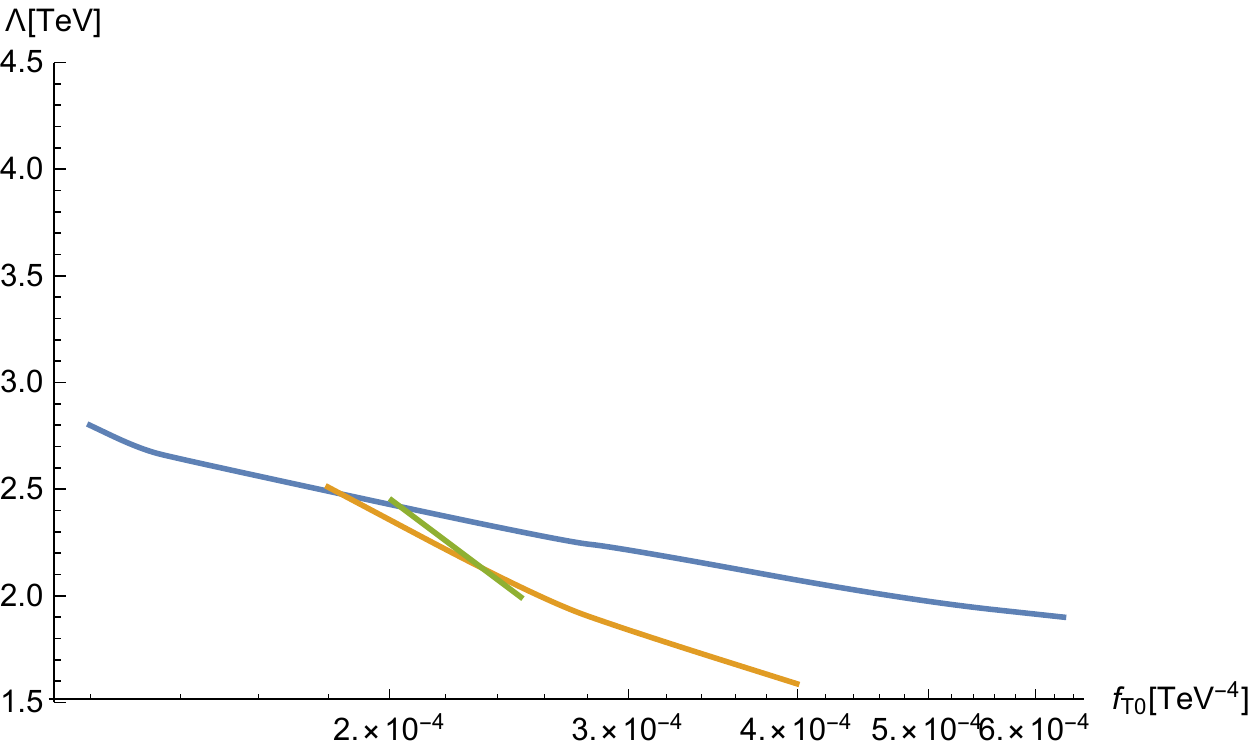}\\ 
\includegraphics[width=0.5\linewidth]{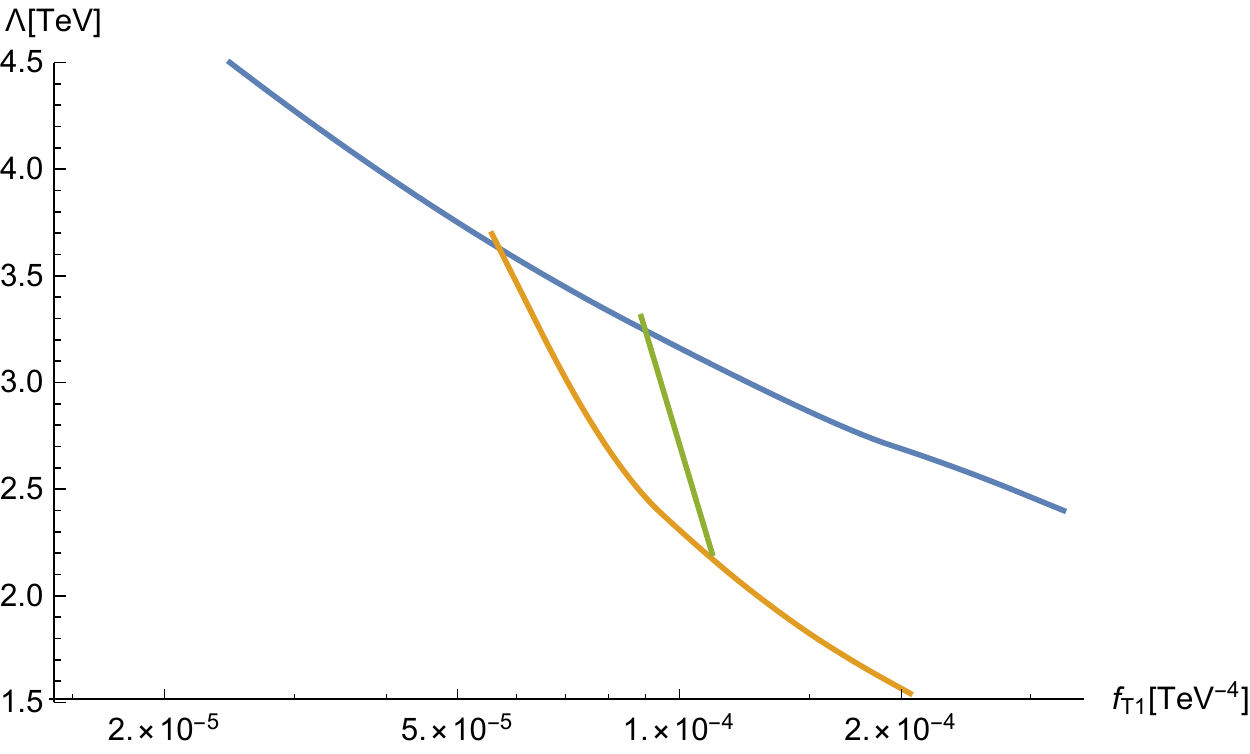} 
&\includegraphics[width=0.5\linewidth]{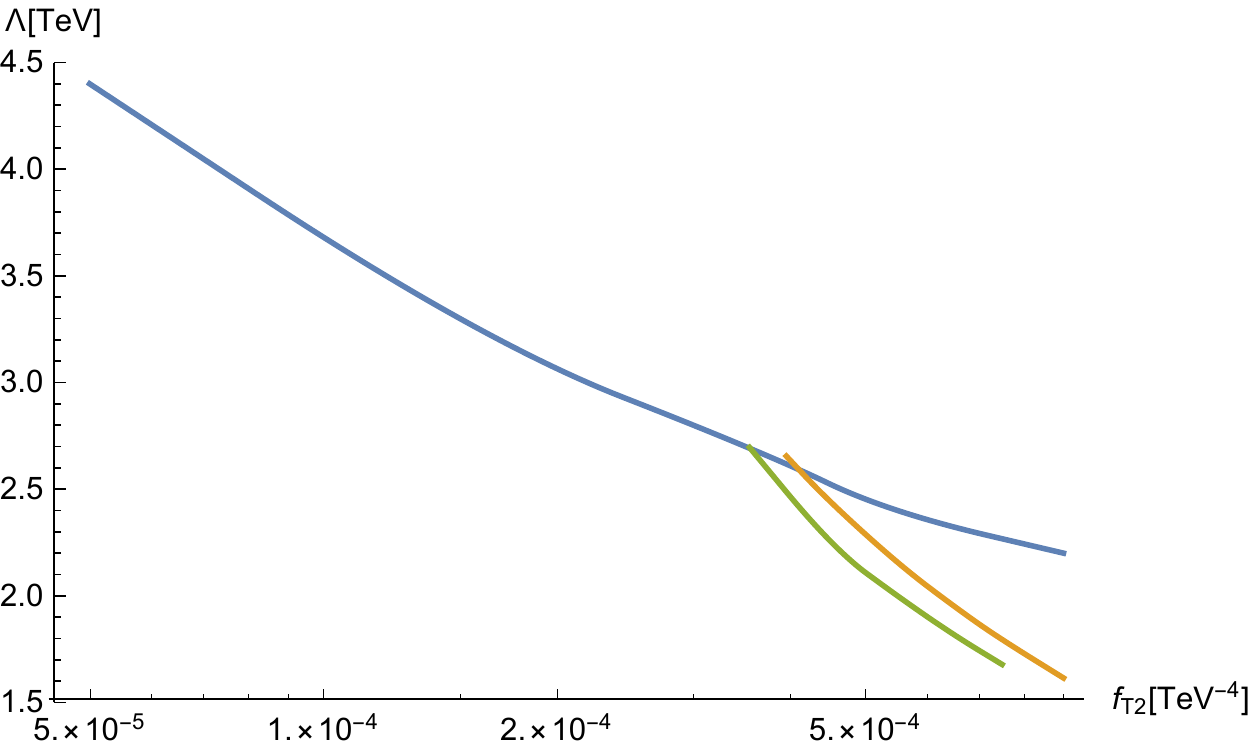}
\end{tabular}
\caption{\em Same description as in Fig.~\ref{fig:trianglesPositivePart1}. From top to bottom and from left to right, the operators considered are $\cT_{62}$, $\cO_{T_1}$, $\cO_{T_2}$ and $\cO_{T_2}$.}
\label{fig:trianglesPositivePart2}
\end{figure}

\begin{figure}[h!] 
\begin{tabular}{cc}
\includegraphics[width=0.5\linewidth]{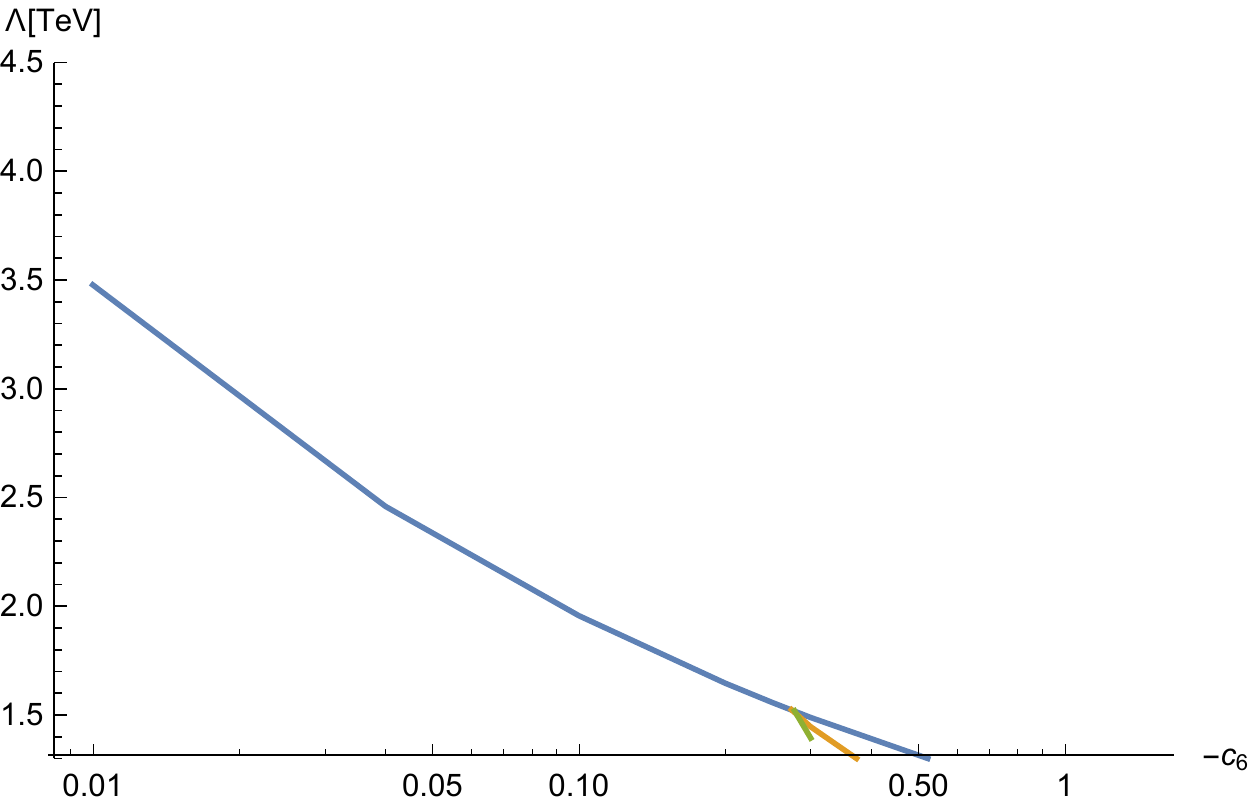} 
& \includegraphics[width=0.5\linewidth]{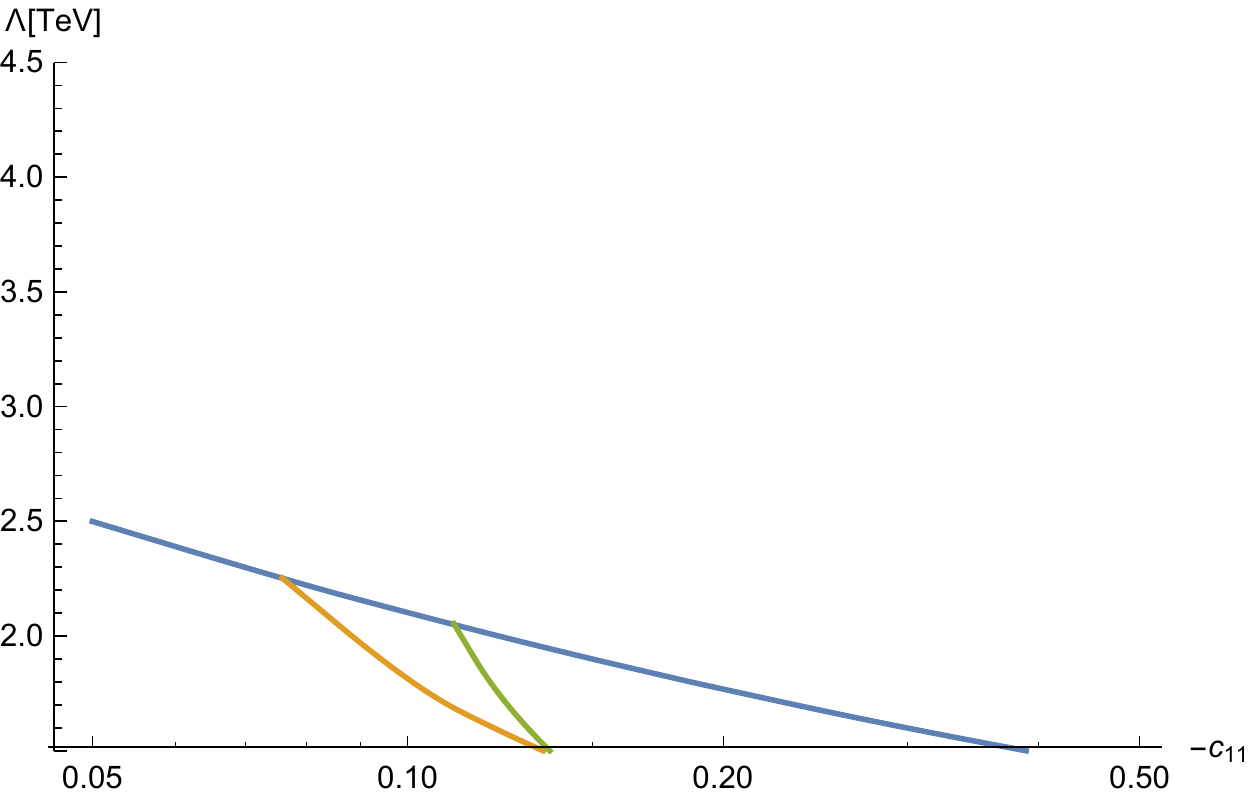} \\
\includegraphics[width=0.5\linewidth]{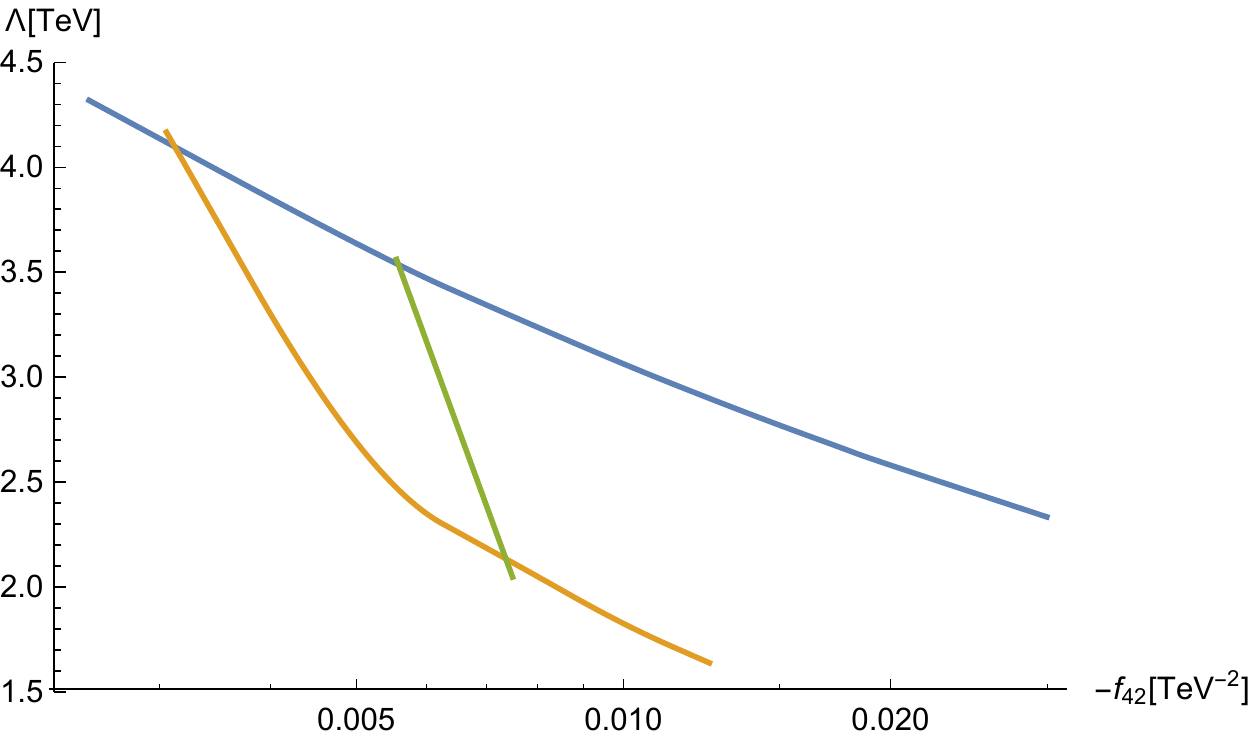} 
& \includegraphics[width=0.5\linewidth]{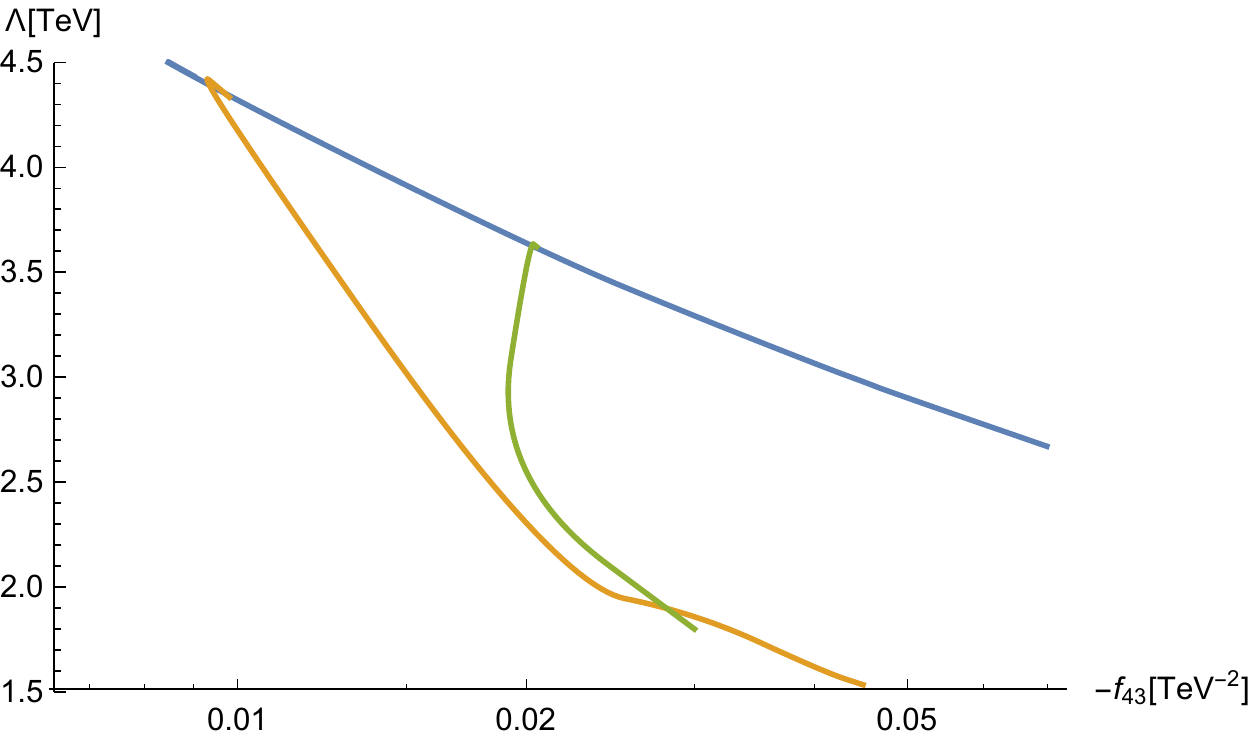} \\
\includegraphics[width=0.5\linewidth]{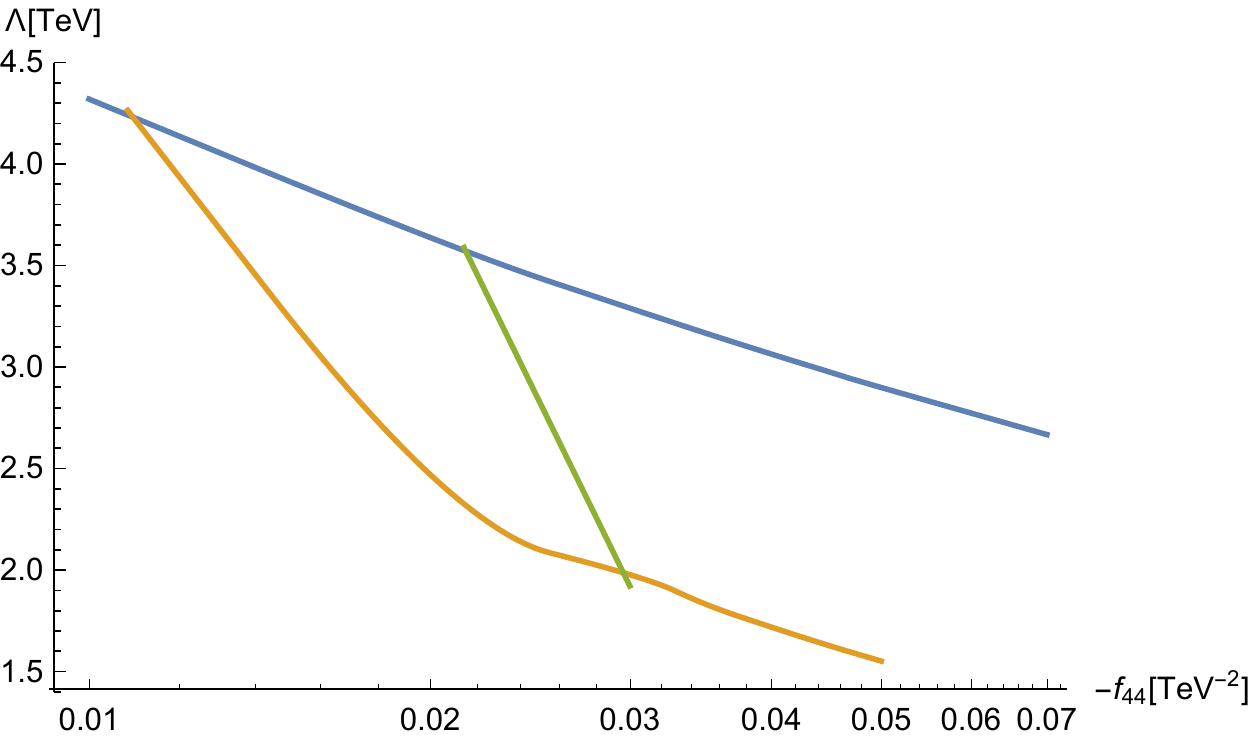} 
&\includegraphics[width=0.5\linewidth]{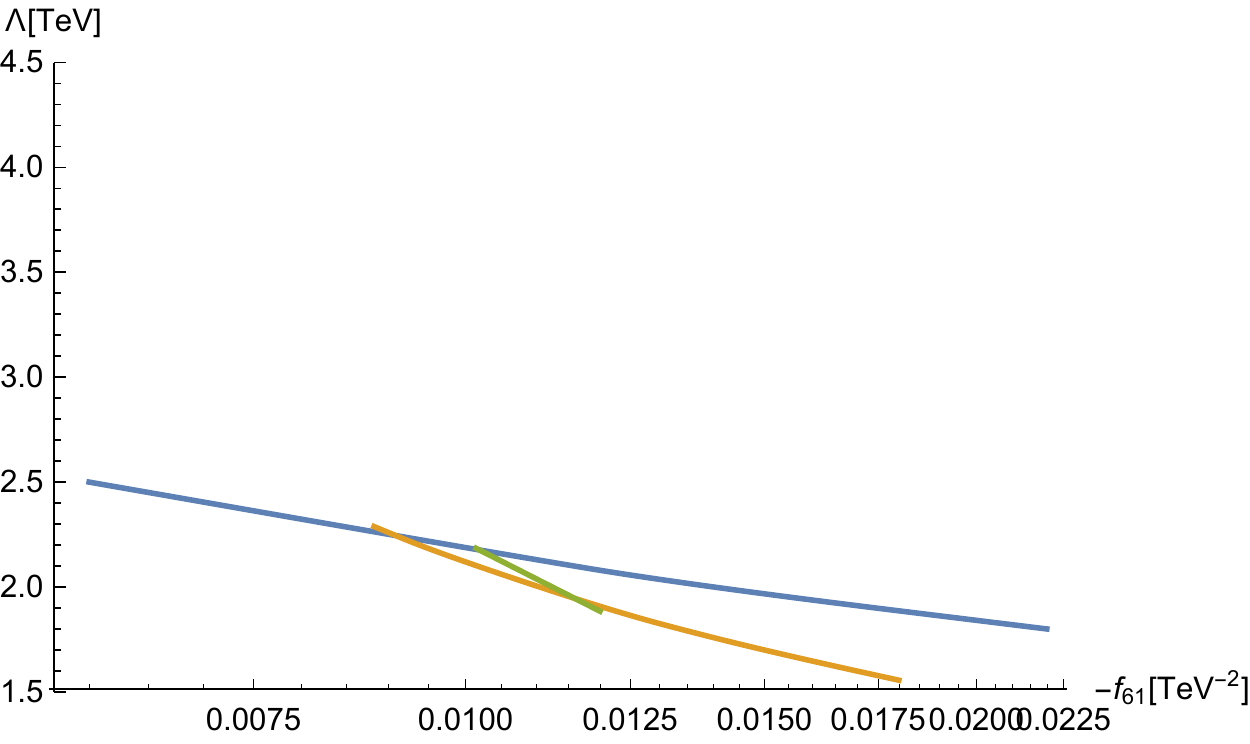} 
\end{tabular}
\caption{\em Same description and operators as in Fig.~\ref{fig:trianglesPositivePart1}. Negative values for $c_i$ and $f_i$ are considered.}
\label{fig:trianglesNegativePart1}
\end{figure}

\begin{figure}[h!] 
\begin{tabular}{cc}
\includegraphics[width=0.5\linewidth]{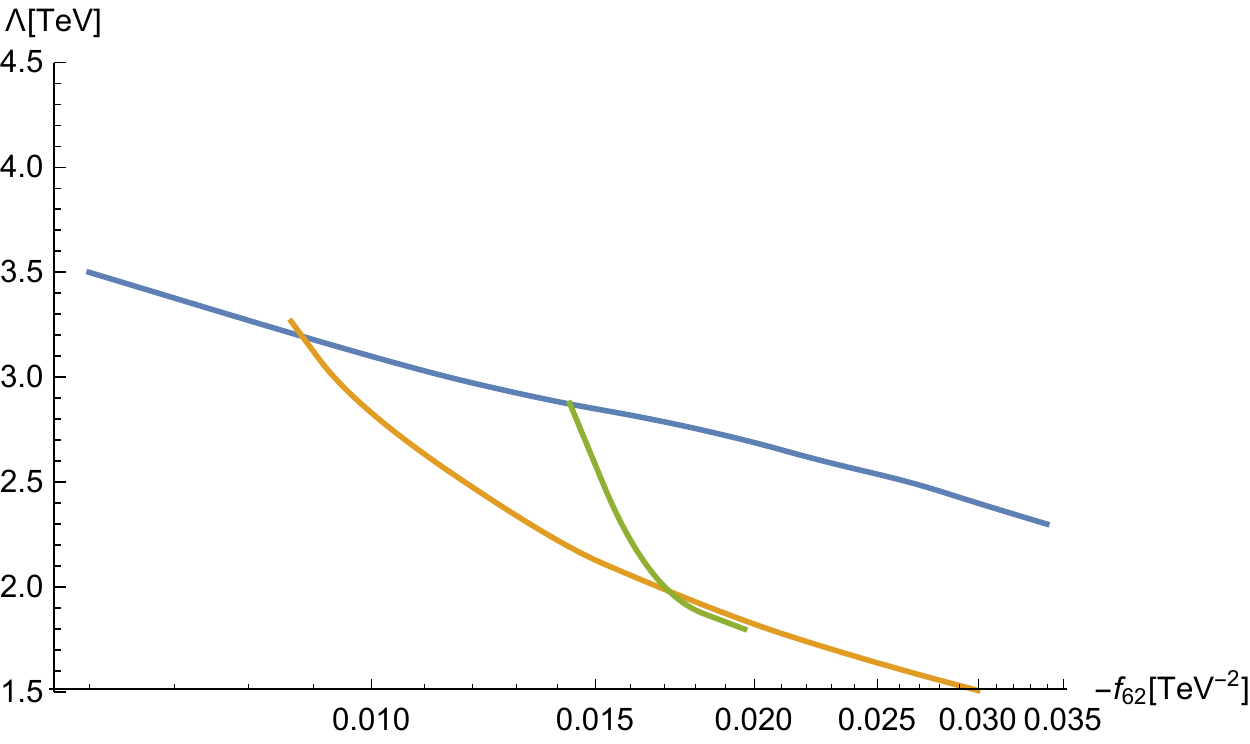}  
& \includegraphics[width=0.5\linewidth]{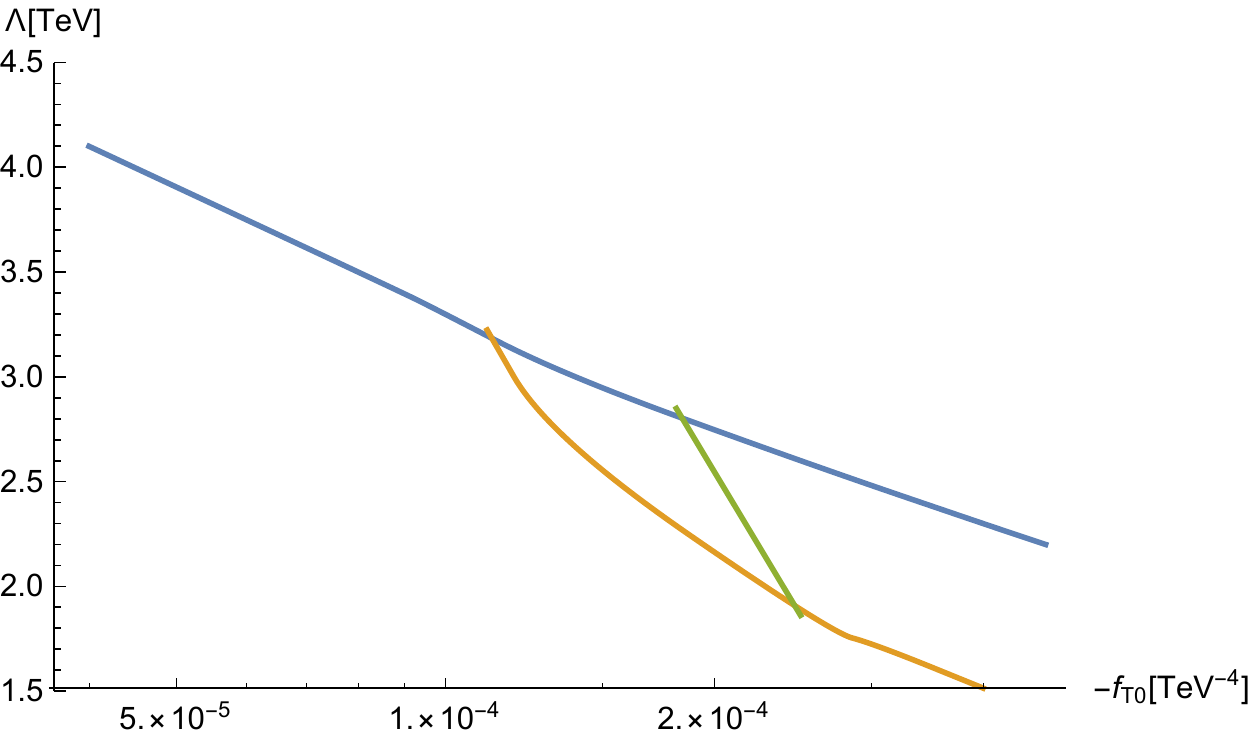}\\
\includegraphics[width=0.5\linewidth]{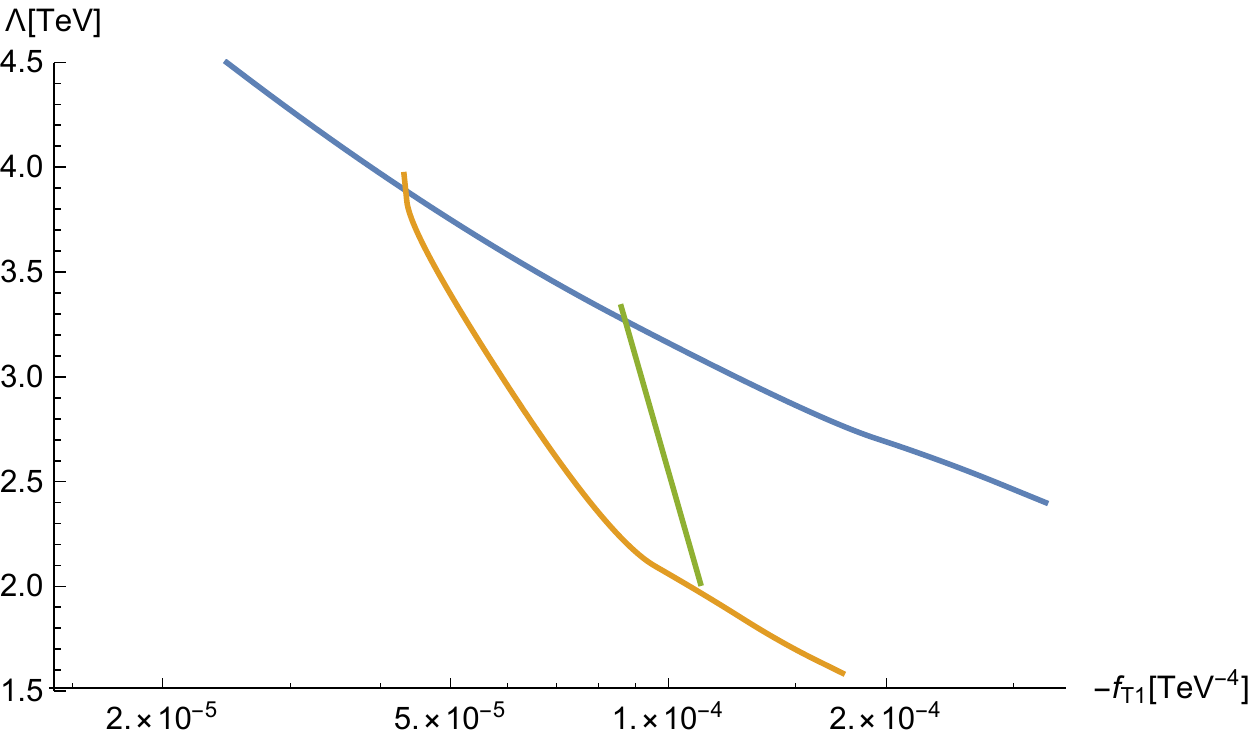} 
& \includegraphics[width=0.5\linewidth]{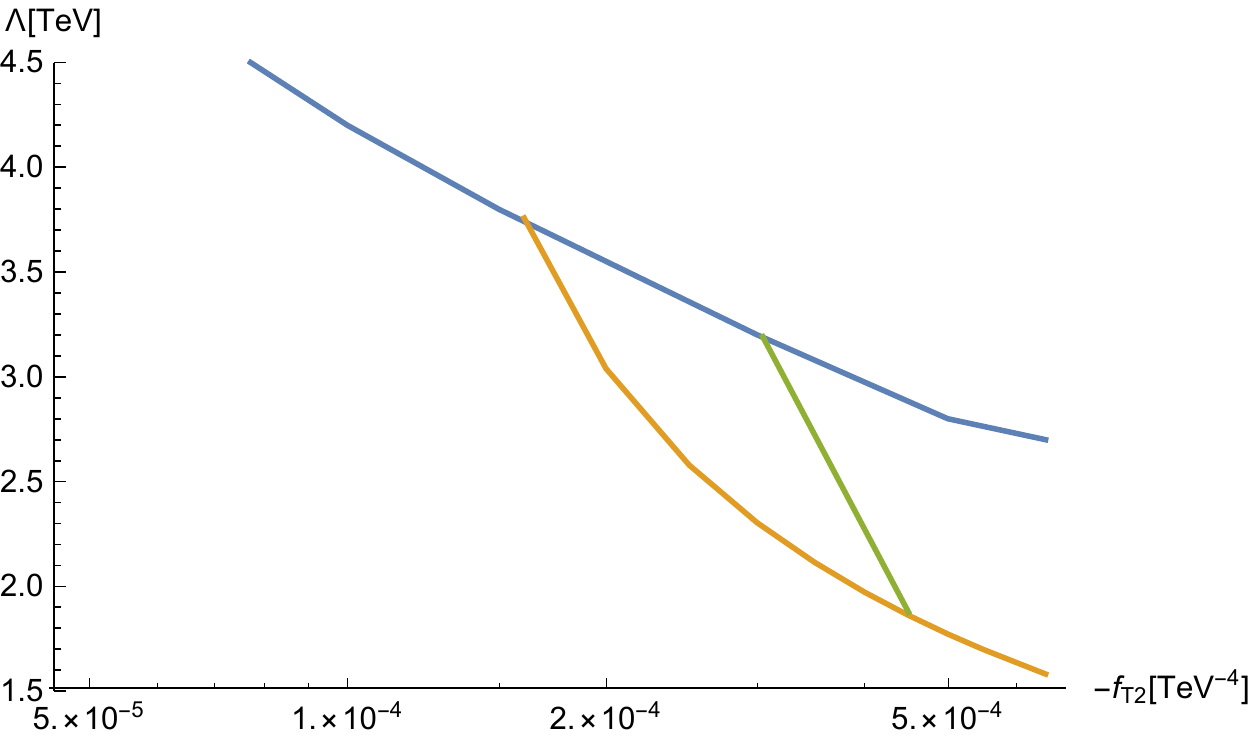}
\end{tabular}
\caption{\em Same description and operators as in Fig.~\ref{fig:trianglesPositivePart2}. Negative values for $c_i$ and $f_i$ are considered.}
\label{fig:trianglesNegativePart2}
\end{figure}

All in all, Figs.~\ref{fig:trianglesPositivePart1} and \ref{fig:trianglesPositivePart2} show the results on the discovery regions for the individual operators $\cP_{6}, \cP_{11},\cT_{42},\cT_{43},\cT_{44}, \cT_{61}, \cT_{62}, \cO_{0},\cO_{1}, \cO_{2}$ for positive $f_i$. The case of $f_i<0$ is shown in Figs.~\ref{fig:trianglesNegativePart1} and ~\ref{fig:trianglesNegativePart2}. The obtained regions resemble triangles in which the left side (yellow line) is bounded by the $5\sigma$ discoverability, the upper side (blue line) by the unitarity violation limit, while the right side (green line) by the 2$\sigma$-EFT consistency.

For the $\cP_6$ and $\cP_{11}$ plots, the vertical axis represents the cut-off scale $\Lambda$ (Eq.~\eqref{eq:pk}) bounded by the energy $M^U$ at which unitarity is violated. 

The discovery region for $\cP_{6}$, in the specific case of a negative Wilson coefficient (Fig.~\ref{fig:trianglesNegativePart1} up-left) is tiny, pointing our a very narrow range of values for $c_6$, centred in $c_6\approx-0.3$, for which a signal could be seen. For a positive Wilson coefficient (Fig.~\ref{fig:trianglesPositivePart1} up-left), there is not any available region, as the discoverability-line is on the right of the $2\sigma$-consistency line: in the region where data may point out a signal, the EFT description breaks down. For this second case, with positive $c_6$, no effects are expected considering the operator $\cP_6$. 

The situation for $\cP_{11}$ (Figs.~\ref{fig:trianglesPositivePart1} and \ref{fig:trianglesNegativePart1} up-right) is different. While for positive Wilson coefficient, the discoverability region is tiny, with $c_{11}\sim0.11$, for a negative Wilson coefficient this region is much larger, with $c_{11}\in-[0.076,\,0.14]$.

These results for $\cP_6$ and $\cP_{11}$ are interesting because they show the impact of the $2\sigma$ EFT consistency requirement considered here for the first time in the HEFT context: in the literature, the coefficients $c_6$ and $c_{11}$ (typically labelled $a_5$ and $a_4$, respectively) may vary within a large range of values providing hypothetically visible signals at colliders; here, instead, chances to find a signal of NP in the $W_LW_L\to W_LW_L$ scattering, described in a consistent HEFT framework by the operators $\cP_6$ and $\cP_{11}$, are present essentially only for negative $c_{11}$, or extremely tuned values of $c_6$ and positive $c_{11}$.

Other cases where the discoverability region is tiny are for $\cT_{61}$ for both signs of the corresponding Wilson coefficients and for $\cO_{T_0}$ for $c_{T_0}>0$, as can be seen in Figs.~\ref{fig:trianglesPositivePart1} and \ref{fig:trianglesNegativePart1} lower-right and Fig.~\ref{fig:trianglesPositivePart2} upper-right, respectively. These regions are centred in $[f_{61}\sim\pm0.01\TeV^{-2},\,\Lambda\sim2.2\TeV]$ for $\cT_{61}$ and $[f_{T_0}\sim2\times 10^{-4}\TeV^{-4},\,\Lambda\sim2.4\TeV]$ for $\cO_{T_0}$, corresponding to $c_{61}\sim\pm0.046$ and $c_{T_0}\sim0.006$. On the other side, another case where no discoverability region is found is for $\cO_{T_2}$ and for positive Wilson coefficient.

\begin{table}[h!]
\hspace{-0.6cm}
\begin{tabular}{|c|c|c|c|c|c|}
\hline
&&&&&\\[-4mm]
   		& $\cP_6$ 	& $\cP_{11}$ 		& $\cT_{42}$ 		& $\cT_{43}$ 		& $\cT_{44}$ 		\\[1mm] 
\hline
&&&&&\\[-3mm]
$c_i>0$ 	& - 		& $0.11$ 			& $[0.033,\,0.007]$ 	& $[0.11,\,0.27]$ 	& $[0.13,\,0.27]$ 	\\[1mm] 
$c_i<0$ 	& $0.3$	& $-[0.076,\,0.14]$ 	& $-[0.034,\,0.070]$	& $-[0.11,\,0.27]$	& $-[0.11,\,0.28]$	\\[1mm]
\hline
\hline
&&&&&\\[-4mm]
  		& $\cT_{61}$ 		& $\cT_{62}$ 		& $\cO_{T_0}$ 			& $\cO_{T_1}$ 			& $\cO_{T2}$ \\[1mm] 
\hline
&&&&&\\[-3mm]
$c_i>0$  	& $[0.045,\,0.047]$  	& $[0.083,\,0.120]$	& $[0.0051,\,0.0072]$ 	& $[0.0026,\,0.0110]$	& - \\[1mm] 
$c_i<0$  	& $-[0.044,\,0.048]$ 	& $-[0.072,\,0.12]$ 	& $-[0.003,\,0.012]$ 		& $-[0.0018,\,0.0110]$ 	& $-[0.0052,\,0.032]$ \\[1mm] 
\hline
\end{tabular}
\caption{\em Ranges of values for the dimensionless $c_i$ operator coefficients corresponding to the discovery regions in Figs.~\ref{fig:trianglesPositivePart1}-\ref{fig:trianglesNegativePart2}. The normalisation is defined in Eq.~\eqref{DeltaLHEFT}. ``-'' denotes no available discovery region.}
\label{tab:cminCmax}
\end{table}

For all the other cases, there are relatively large discovery regions where the corresponding operators enhance the signal with respect to the SM prediction, with a possibility for this enhancement to be measured at colliders and with a consistent EFT description. The Tab.~\ref{tab:cminCmax} shows the ranges of values for the dimensionless coefficients $c_i$, with the normalisation as in Eq.~\eqref{DeltaLHEFT}, for all the discovery regions in Figs.~\ref{fig:trianglesPositivePart1}-\ref{fig:trianglesNegativePart2}. As can be seen, all the coefficients are smaller than 1 and, considering that the NDA normalisation has been adopted in Eq.~\eqref{DeltaLHEFT}, this leads to the conclusion that the contributions considered here are the dominant ones: in other words, higher order quantum corrections to the operator list in Eq.~\eqref{HEFTOps} are subdominant.

As pointed out in Sect.~\ref{sec:comparison}, a few HEFT operators considered in the previous analysis can find a sibling in the SMEFT at $d=8$, that have been the focus of the study in Ref.~\cite{Kalinowski:2018oxd}. The discoverability regions show in 
Figs.~\ref{fig:trianglesPositivePart1}-\ref{fig:trianglesNegativePart2} for these operators indeed match with the results presented in Ref.~\cite{Kalinowski:2018oxd}, after the rescaling in order to account for the normalisations. This can be easily checked for the operators $\cO_{T_1}$ that are the same in both the bases and the operator $\cP_6$ that is equivalent to $\cO_{S_1}$; it is slightly more difficult for the operator $\cP_{11}$ that is equivalent to a combination between $\cO_{S_0}$ and $\cO_{S_1}$, as explicitly shown in Eq.~\eqref{eq:correlations}. The small differences occur for small values of $\Lambda$, where there is less available discoverability region, see e.g. for $\cO_{T_0}$, below $\Lambda\sim 2.5\TeV$. They are due to different (updated) software tools and different analysis algorithms within these tools. On the other side, these differences may represent the uncertainties on the discovery region determination.

The last three operators, $\cT_{42}$, $\cT_{43}$ and $\cT_{44}$, do no have a sibling in the SMEFT Lagrangian at $d\leq8$. The interactions described by $\cT_{42}$, $\cT_{43}$ and $\cT_{44}$ will be described in terms of $d>8$ SMEFT operators, and therefore the strength of their signal is expected to be much more suppressed. Moreover, the helicity amplitudes that are considerably enhanced or even dominating the cross section at $M_{WW}\lesssim\Lambda$ are different than those found for the SMEFT operators (for the total unpolarised cross sections and its polarised fractions in on-shell $WW$ scattering with $\cT_{42}$, $\cT_{43}$ and $\cT_{44}$ insertions see the Appendix; see also Appendix in Ref.~\cite{ww27paper} for the SMEFT case). This suggests that the same-sign $WW$ scattering could be a sensitive channel to disentangle between the SMEFT and HEFT descriptions of the EWSB sector, but a more dedicated analysis would be necessary to investigate this possibility.\\

The results presented so far for the HEFT Lagrangian can be translated in terms of more fundamental theories whose dynamics takes place at a much higher energy scale. The first UV example is the one of generic CH models as described in Refs.~\cite{Alonso:2014wta,Hierro:2015nna}: instead of considering the three SM GBs and the physical Higgs as independent objects as in the HEFT, Refs.~\cite{Alonso:2014wta,Hierro:2015nna} treated the four fields as a doublet of the EW symmetry and presented the effective chiral Lagrangian for a generic symmetric coset. For the specific case of $SO(5)/SO(4)$, the initial Lagrangian is written invariant under the global $SO(5)$ symmetry; after the spontaneous breaking down to $SO(4)$, the SM GBs and the physical Higgs arise as GBs of the symmetric coset, as  a bi-doublet of $SU(2)_L\times SU(2)_R\equiv SO(4)$ GBs, described altogether by a single unitary matrix. Effective chiral operators can be constructed on the same line as in the EW$\chi$L. After the explicit $SO(5)$ symmetry breaking, the physical Higgs becomes massive and can be distinguished from the other GBs. After this breaking, the Lagrangian describing the low-energy model matches the HEFT one, where the Wilson coefficients $c_i$ are written in terms of the high-energy coefficients, $\tilde c_i$ in the notation of Refs.~\cite{Alonso:2014wta,Hierro:2015nna}. 

The analysis in Refs.~\cite{Alonso:2014wta,Hierro:2015nna} considered up to four-derivative operators and therefore, once focusing only on the genuine quartic operators, the results of Tab.~\ref{tab:cminCmax} in this paper can be used to constrain the operator coefficients $\tilde c_4$, $\tilde c_5$ and $\tilde c_6$ that appear in Tab.~1 in Ref.~\cite{Alonso:2014wta}. Adopting the NDA normalisation, the $c_6$ and $c_{11}$ coefficients can be written as
\be
c_6=-8\,\pi^2\,\xi\,\tilde c_6+16\,\pi^2\,\xi^2\,\tilde c_4\,,\qquad\qquad
c_{11}=16\,\pi^2\,\xi^2\,\tilde c_5\,.
\ee
An explicit value for $\xi$ can be extracted once taking the values in Tab.~\ref{tab:cminCmax}: assuming, for example, the specific values for $\tilde c_i$ coefficients indicated inside the brackets,
\be
\begin{aligned}
c_6(\tilde c_4=-1,\,\tilde c_6=0)&=-0.3\quad&&\Longrightarrow\quad \xi=0.04\\
c_6(\tilde c_4=0,\,\tilde c_6=1)&=-0.3\quad&&\Longrightarrow\quad \xi=0.004\\
c_{11}(\tilde c_5=1)&=0.11\quad&&\Longrightarrow\quad \xi=0.026\\
c_{11}(\tilde c_5=-1)&=-[0.076,\,0.14]\quad&&\Longrightarrow\quad \xi=[0.026,\,0.03]\,.
\end{aligned}
\label{XiValuesfromCoefficients}
\ee

As a second example, one can consider the so-called Minimal Linear $\sigma$ Model~\cite{Feruglio:2016zvt}, that is a renormalisable model which may represent the UV completion of the Minimal CH model. Also in this case, only up to four-derivative operators have been considered in the low-energy limit; while the Wilson coefficient $c_{11}$ does not receive any contribution, $c_6$ turns out to be dependent only from $\tilde c_4=1/64$. The parameter $\xi$ is fixed to be
\be
\xi=0.35\,.
\label{xiFromP6}
\ee
This result should be interpreted as follows: in case of a NP discovery in the $W^+W^+$ scattering with a significance of more than $5\sigma$, in the Lorentz configurations described by the $\cP_6$ operator, and assuming that this NP corresponds to the Minimal Linear $\sigma$ Model, the parameter $\xi$ would take the value in Eq.~\eqref{xiFromP6}. However, such a large value would be already excluded considering EW precision observable constraints~\cite{Feruglio:2016zvt} (for a review see Ref.~\cite{Panico:2015jxa}). It follows that the Minimal Linear $\sigma$ Model cannot explain such a NP signal on $W^+W^+$ scattering.

An alternative possibility to extract bounds on $\xi$ is to consider the traditional relation between the scale $f$ and the cut-off $\Lambda$ present in the CH scenario~\cite{Kaplan:1983fs},
\be
f<\Lambda<4\pi f\,.
\ee
After a simple manipulation, this expression can be written in terms of the parameter $\xi$ as
\be
\dfrac{v^2}{\Lambda^2}<\xi<16\pi^2\dfrac{v^2}{\Lambda^2}\,.
\ee
This relation provides model independent bounds on $\xi$ that can be derived looking at the discoverability plots in Figs.~\ref{fig:trianglesPositivePart1}-\ref{fig:trianglesNegativePart2}. Considering again, as an example, the $\cP_{11}$ operator, 
\be
\begin{cases}
c_{11}>0\quad\rightarrow\quad\Lambda\approx2\TeV\quad&\Longrightarrow\quad 0.015<\xi<2.4\\
c_{11}<0\quad\rightarrow\quad1.5\TeV\lesssim\Lambda\lesssim2.3\TeV\quad&\Longrightarrow\quad 0.011<\xi<4.2\\
\end{cases}
\ee
compatible with the results in Eqs.~\eqref{XiValuesfromCoefficients} and \eqref{xiFromP6}. Similar ranges of values can be found for the other operators.

\boldmath
\section{Conclusions}
\label{Sect.Conclusions}
\unboldmath

This paper presents the analysis of the prospects for discovering new physics at the HL-LHC considering the process $pp\rightarrow W^+ W^+jj$ within the HEFT framework. The focus is on the same-sign $WW$ scattering with purely leptonic $W$ decays, described by a set of effective operators that encode the new physics contributions. Only genuine quartic operators up to $d_p=8$ defined in Eq.~\eqref{HEFTOps} have been considered, that is operators contributing to quartic gauge couplings but not to triple gauge couplings.

The analysis follows the strategy illustrated in Ref.~\cite{Kalinowski:2018oxd} for the SMEFT and consists in identifying the region of the parameter space that satisfies  three conditions: for each of the operators, i) the signal estimate $D_i^\text{BSM}$ should deviate from the SM prediction for at least $5\sigma$; ii) the c.o.m. energy scale $M_{WW}$ of the process should be below the cut-off scale $\Lambda$, that should be below the scale $M^U$ at which unitarity gets violated; this means that the only relevant data in an EFT fit must belong to the region where $M_{WW}<\Lambda$; in particular the tail of the $M_{WW}$ distribution with $M_{WW}>\Lambda$ should not have any impact on the fit. This requirement is implemented quantitatively by the third condition: iii)
one introduces two signal estimates, one coming uniquely from the EFT in its range of validity and assuming only the SM contribution in the region where $\Lambda<M_{WW}<M_{WW}^\text{max}$ and the second estimate, with a tail regularisation of the distribution in $M_{WW}$.
In order to guarantee that also with this second signal estimate the physical effects in the fit do not come from the tail of the distribution, statistical consistency between the two estimates within two standard deviations ($2\sigma$) is required. 
The first condition implies that evidence of NP occurs in the same-sine $WW$ process, while the second guarantees that the EFT description is consistently adopted in the analysis, and finally the last condition assures that the physical effects in the fit do not come from the tail of the distribution, where the effective theory description is not valid anymore. The discoverability regions, when non-vanishing, are similar to triangles from which it is possible to extract bounds on the operator coefficients and, in case, on the cut-off of the theory.

These results can be translated into bounds on the parameters of UV theories that project at low-energy on the HEFT operators. The case of the CH model has been discussed, considering first a generic $SO(5)/SO(4)$ CH description and then the so-called Minimal Linear $\sigma$ Model. In both cases, interesting bounds have been found on the parameter $\xi$, that measures the level of non-linearity of the Higgs sector and its fine-tuning. These bounds are independent from those extracted considering EW precision observables or the absence of any composite resonance.

When comparing the results presented here with the past literature, it is worth mentioning that in Ref.~\cite{Delgado:2013hxa} values for the coefficients $a_4$ and $a_5$ equal to $0.005$ have been considered. These coefficients correspond to $c_{11}$ and $c_6$, respectively, in the notation used here. Translating that value in the NDA normalisation adopted in this paper, it turns out to be equal to 0.8, and therefore it is outside the discoverability regions identified for the operators $\cP_{6}$ and $\cP_{11}$. The reason of this discrepancy is the additional constraint, considered here for the first time, of the EFT validity, that is conditions ii) and iii) listed above. This exemplifies the relevance of the procedure illustrated here to interpret data on same-sign $WW$ scattering in terms of the HEFT Lagrangian.

\section*{Acknowledgements}

The work of P.K. is supported by the Spanish MINECO project FPA2016-78220-C3-1- P (Fondos FEDER) and by National Science Centre, Poland, the PRELUDIUM project under contract 2018/29/N/ST2/01153. L.M. acknowledges partial financial support by the Spanish MINECO through the ``Ram\'on y Cajal'' programme (RYC-2015-17173), by the Spanish ``Agencia Estatal de Investigaci\'on'' (AEI) and the EU ``Fondo Europeo de Desarrollo Regional'' (FEDER) through the project FPA2016-78645-P, and through the Centro de excelencia Severo Ochoa Program under grant SEV-2016-0597, and by the European Union's Horizon 2020 research and innovation programme under the Marie Sklodowska-Curie grant agreements No 690575 and No 674896. The work of S.P.  is partially supported by the National Science Centre, Poland, under research grants DEC-2015/18/M/ST2/00054 and DEC-2016/23/G/ST2/04301. M.S. is partially supported by the generous COST grant, COST Action No. CA16108 (VBSCan).

L.M. thanks the Institute of Theoretical Physics of the University of Warsaw for hospitality during the development of this project. S.P. thanks the Instituto de F\'isica Te\'orica (IFT UAM-CSIC) in Madrid for its support via the Centro de Excelencia Severo Ochoa Program under Grant SEV-2016-0597.

\appendix

\begin{boldmath}
\section{Details on the Same-Sign $WW$ Scattering}
\label{App.A}
\end{boldmath}

The on-shell same-sign $WW$ scattering total unpolarised cross section and its polarised fractions as functions of $M_{WW}$ are illustrated in this appendix, for chosen values of $f_i=c_i/\Lambda^{n-4}$. The total unpolarised cross-section reads
\begin{equation}
\sigma\sim\frac{1}{9}\ \sum_{i,j,k,l}\ \ \left|A_\text{SM}(ij\rightarrow kl)\right|^2 + (A_\text{SM}(ij\rightarrow kl)A_\text{BSM}(ij\rightarrow kl)^{\ast} + \hc) + \left|A_\text{BSM}(ij\rightarrow kl)\right|^2,
\label{eq:app1}
\end{equation}
where $A_\text{SM}$, $A_\text{BSM}$ are the SM and BSM parts of the scattering amplitude, $A=A_\text{SM}+A_\text{BSM}$, for $ijkl$ $W$'s polarisations. $A_\text{BSM}$ is the part proportional to the operator coefficients $c_i$. 

In the helicity basis, that will be adopted in the rest of the section, by polarised fractions is meant the single $ijkl$ contribution to Eq.~\eqref{eq:app1}. There are $3^4=81$ such contributions that can be divided into classes that yield the same (polarised) cross sections due to $P$ and $T$ discrete symmetries and Bose statistics. Hence, a reduced number of 13 independent polarisation classes can be considered, taking into account multiplicity factors while computing the cross section (see the appendix in Ref.~\cite{Kalinowski:2018oxd} for details).

\begin{figure}[h!] 
  \begin{tabular}{cc}
      \includegraphics[width=0.5\linewidth]{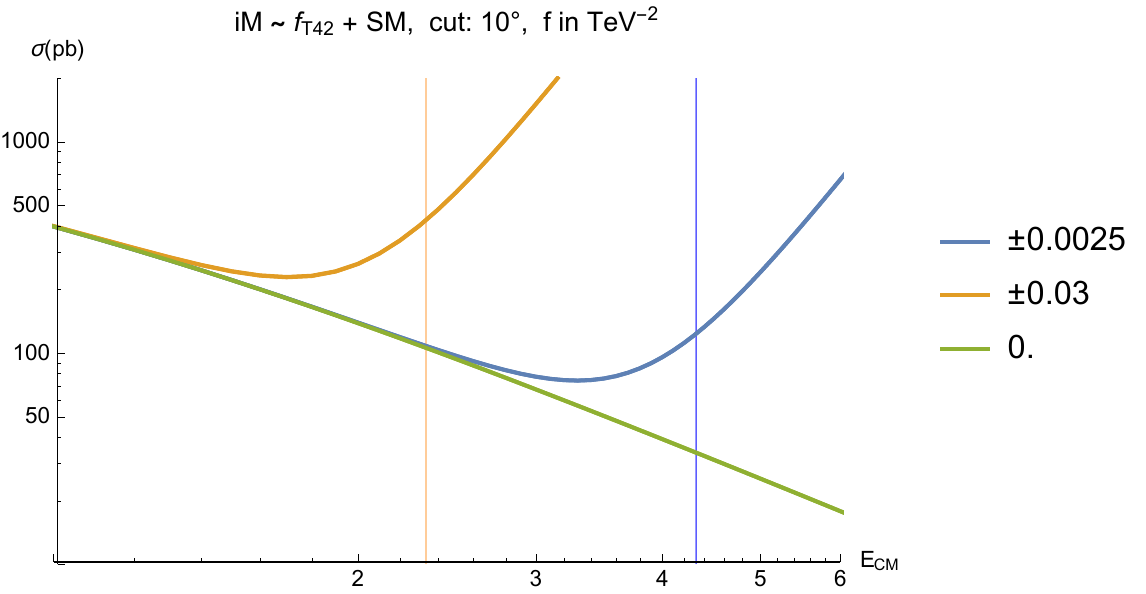} 
& \includegraphics[width=0.5\linewidth]{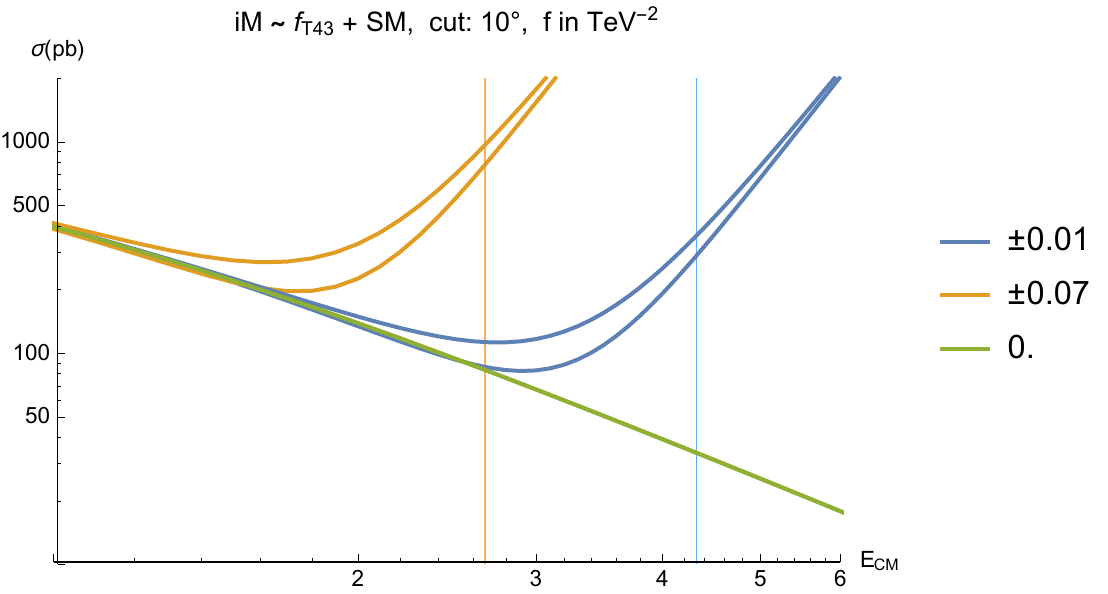} \\
  \end{tabular}
\begin{center}
	\includegraphics[width=0.5\linewidth]{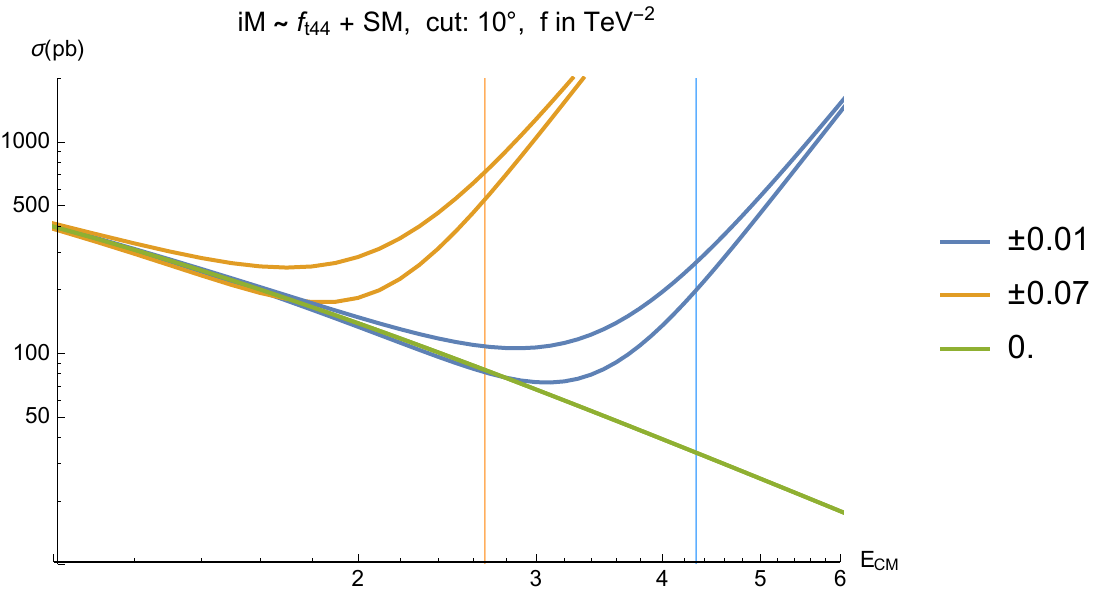}
\end{center}
\caption{\em Energy dependence of the total unpolarized $W^+W^+$ cross sections ($E_{CM} \equiv M_{WW}$, in TeV) for a chosen set of $f_i$ values. Vertical lines denote the unitarity bound $M^U$ (color correspondence). There is no color distinction between the signs: for both $\cT_{43}$ and $\cT_{44}$, the upper cross section curves correspond to $f_i<0$, while the lower ones to $f_i>0$.}
\label{fig:unpol}
\end{figure}

Fig.~\ref{fig:unpol} shows the total unpolarised $WW$ cross sections as functions of $M_{WW}$ for $\cT_{42}$, $\cT_{43}$, and $\cT_{44}$ operators, for both signs of the Wilson coefficients. The vertical lines denote unitarity limits $M^U$. All cross sections are computed with a $10^{\circ}$ cut in the forward and backward scattering regions.

\begin{figure}[h!]
\begin{center}	\includegraphics[width=0.95\textwidth]{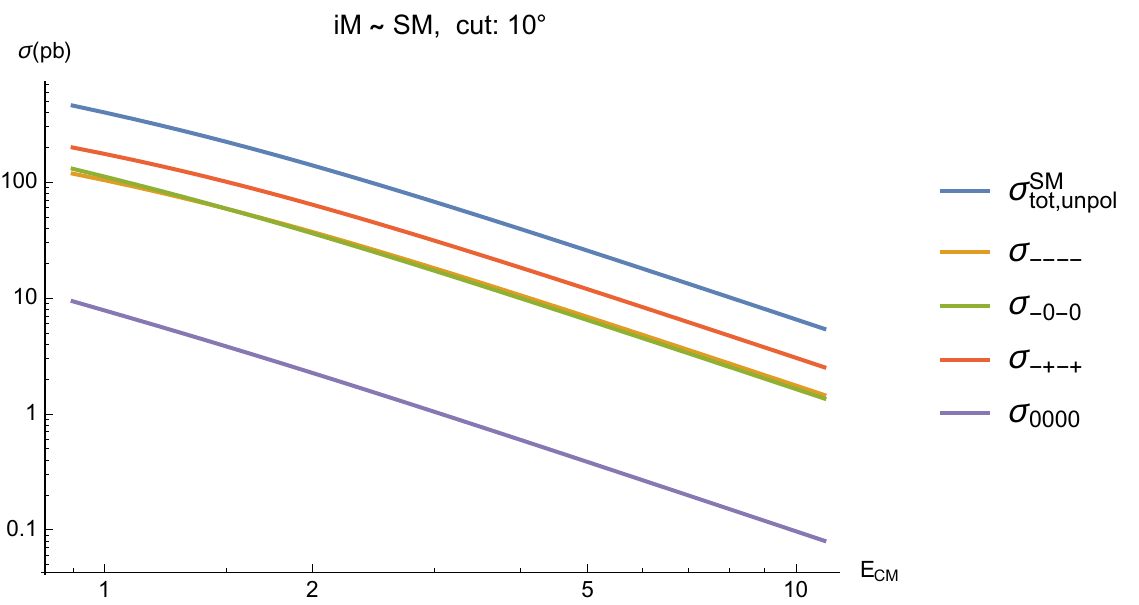}
\caption{\em Contributions of different polarisations (multiplicity taken into account) to the total unpolarised cross section as functions of the c.o.m. $WW$ energy ($E_{CM} \equiv M_{WW}$, in TeV) in the SM. The total unpolarised cross section is also shown in blue.}
\label{fig:polSM}
\end{center}
\end{figure}

\begin{figure}[h!] 
  \begin{tabular}{cc}
      \includegraphics[width=0.5\linewidth]{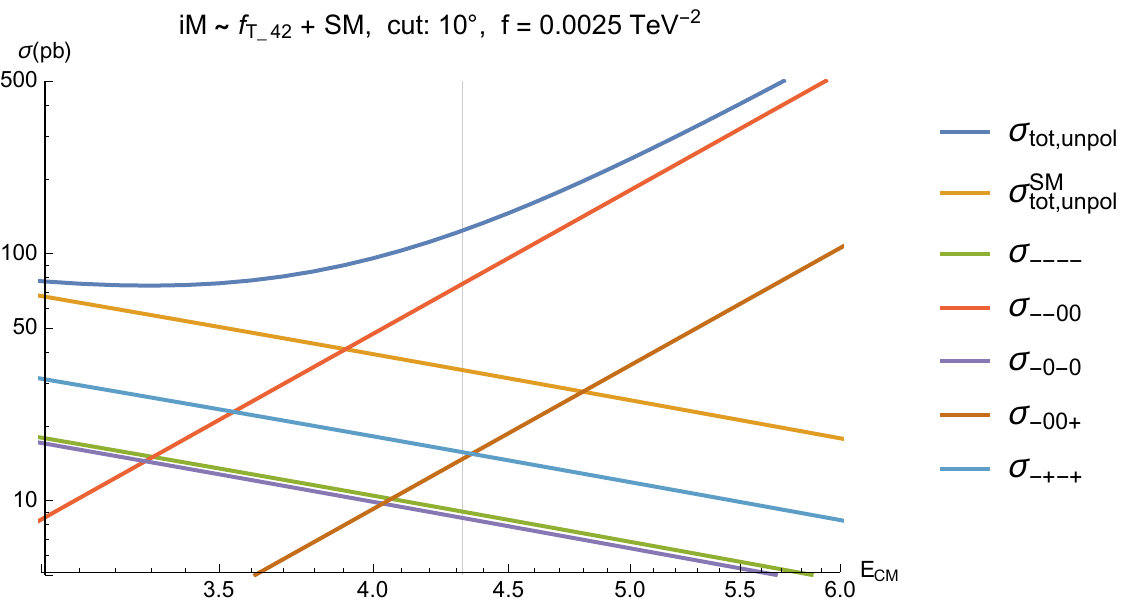} 
& \includegraphics[width=0.5\linewidth]{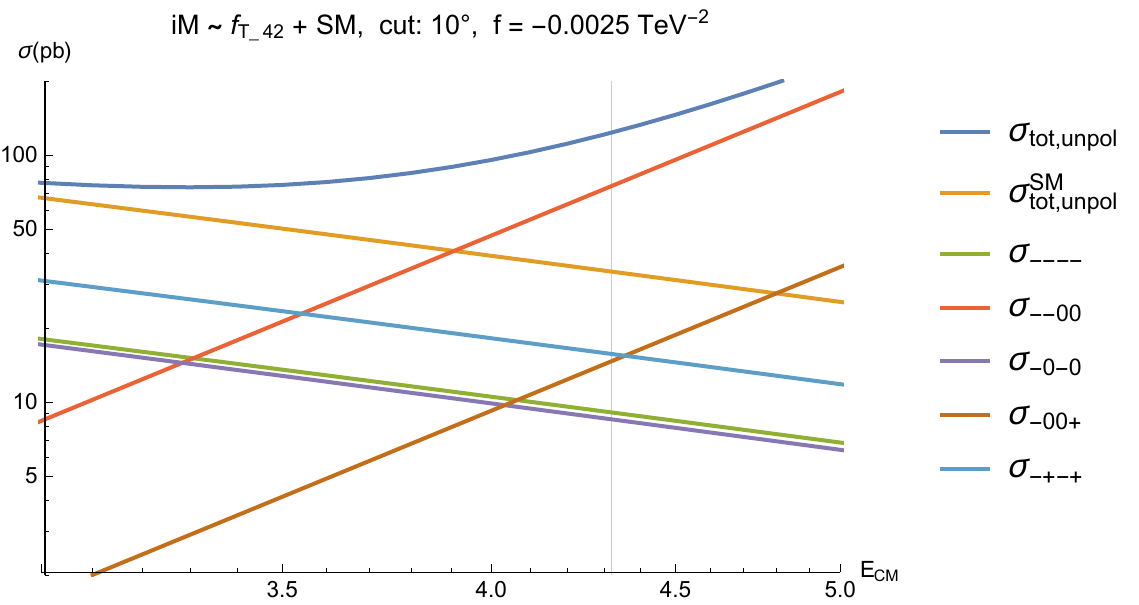} \\
\includegraphics[width=0.5\linewidth]{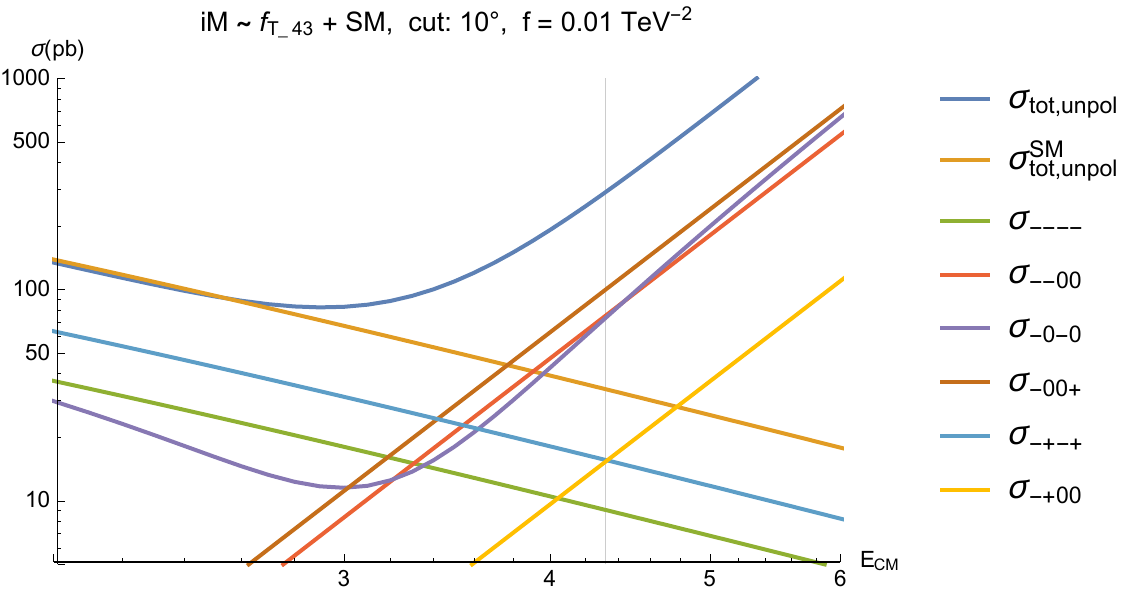} 
& \includegraphics[width=0.5\linewidth]{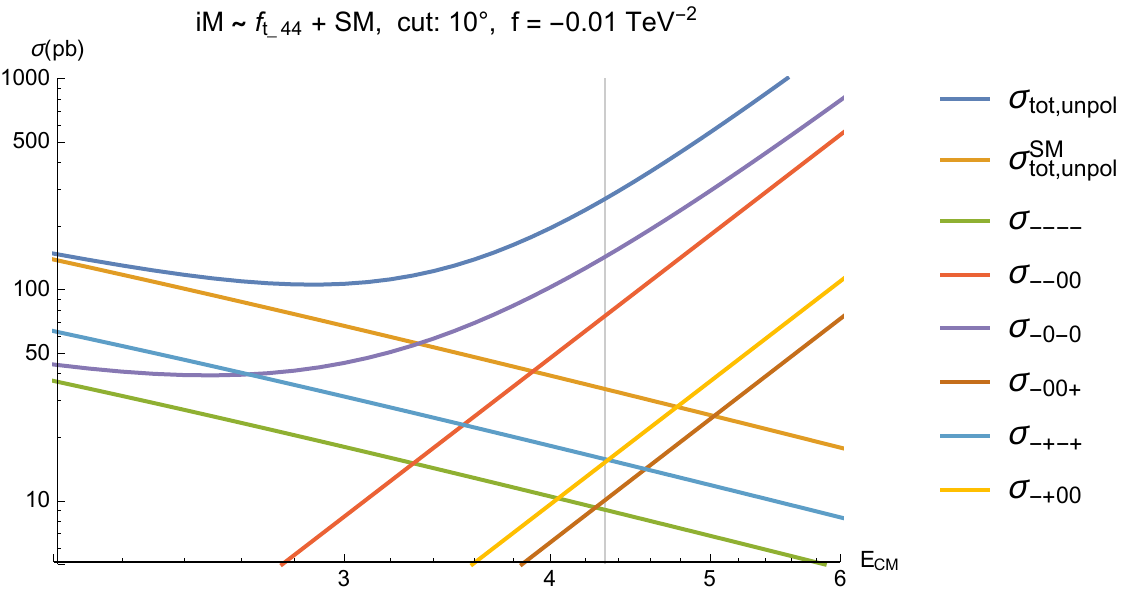} \\
\includegraphics[width=0.5\linewidth]{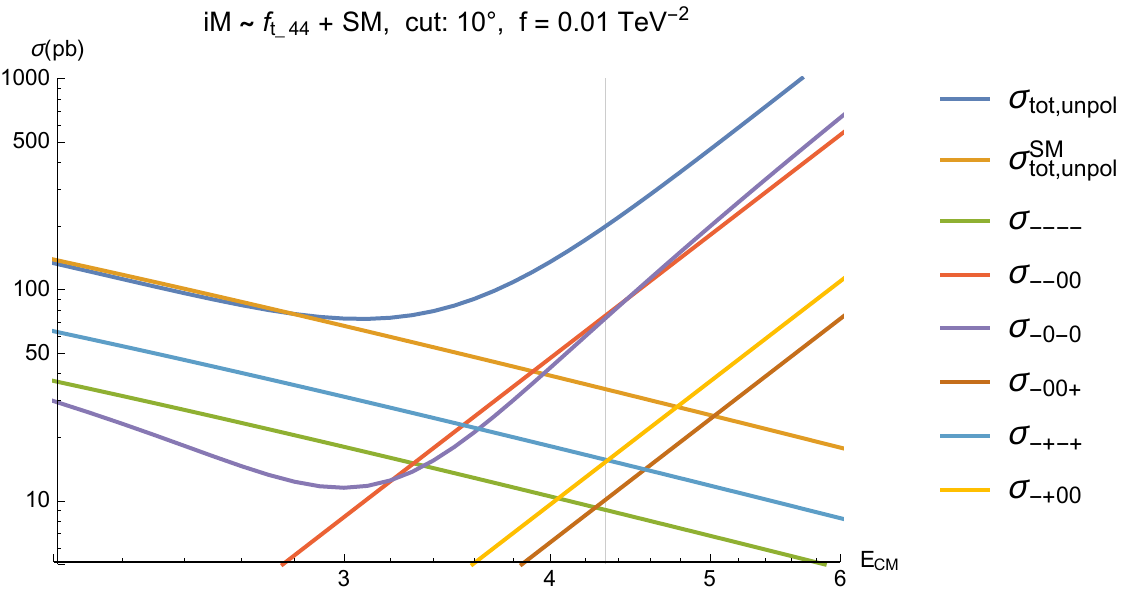} 
& \includegraphics[width=0.5\linewidth]{PLOT/plotlistPlot10DegreeParticPolsft44Eq001LeqZoom.pdf}
  \end{tabular}
\caption{\em The polarised contributions to the total unpolarised cross sections (multiplicity taken into account) as functions of the c.o.m. collision energy ($E_{CM} \equiv M_{WW}$, in TeV) for chosen values of $f_i$. The left column corresponds to $f_i>0$, while the right one to $f_i<0$. The remaining (not shown) polarised cross sections are negligibly small. The total cross sections and the SM total cross sections are also shown.}
\label{fig:pol}
\end{figure}

Fig.~\ref{fig:polSM} shows the polarized cross section fractions for  the SM. Fig.~\ref{fig:pol} shows the polarised cross section fractions for $\cT_{42}$, $\cT_{43}$, and $\cT_{44}$ and for chosen values of $f_i$, separating the two signs of $f_i$ on different plots. Notice however that no sign dependence is present in $\cT_{42}$, due to the fact that the enhanced polarised cross sections $--00$ and $--00+$ are not among the polarised cross sections that saturate in the SM (see Appendix in Ref.~\cite{ww27paper} for details on non-interference in $WW$ scattering in the context of the SMEFT operators).

The enhancement of $-+0+$ and/or $-+00$ fractions in the region $M_{WW}\lesssim\Lambda$, visible in fig.~\ref{fig:pol}, occurs only for the operators $\cT_{42}$, $\cT_{43}$, and $\cT_{44}$ among all the HEFT operators studied in this work. This can be inferred from the appendix in Ref.~\cite{Chaudhary:2019aim} where the genuine quartic $d=8$ SMEFT operators are studied.The remaining HEFT operators obey simple relations with corresponding SMEFT operators (see Sect.~\ref{sec:comparison}).

\pagebreak
\newpage
\footnotesize

\providecommand{\href}[2]{#2}\begingroup\raggedright\endgroup

\end{document}